\documentclass[twocolumn]{aastex631}
\usepackage{xcolor}
\usepackage{amsmath,amssymb}	
\usepackage{subfigure,wrapfig}
\usepackage{xspace}
\usepackage{mathtools}
\DeclarePairedDelimiter\abs{\lvert}{\rvert}

\definecolor{nicegreen}{RGB}{9, 219, 31}

\newcommand{\rstar}{\ensuremath{R_{\star}}}
\newcommand{\bhac}{\texttt{BHAC}\xspace}
\newcommand{\raptor}{\texttt{RAPTOR}\xspace}
\newcommand{\xpsi}{\texttt{X-PSI}\xspace}
\newcommand{\iobs}{\ensuremath{i_{\rm obs}}}

\begin{document}

\title{Pulse Profiles of Accreting Neutron Stars from GRMHD Simulations}

\author[0000-0002-4764-6189]{Pushpita Das}
\affiliation{Anton Pannekoek Institute for Astronomy, University of Amsterdam, Science Park 904, 1098 XH, The Netherlands}
\affiliation{Department of Astronomy, Columbia University, New York, NY 10027, USA}

\correspondingauthor{Pushpita Das}
\email{p.das2@uva.nl}

\author[0000-0001-6356-125X]{Tuomo Salmi}
\affiliation{Anton Pannekoek Institute for Astronomy, University of Amsterdam, Science Park 904, 1098 XH, The Netherlands}
\affil{Department of Physics, P.O. Box 64, FI-00014 University of Helsinki, Finland}


\author[0000-0002-2685-2434]{Jordy Davelaar}
\affiliation{Department of Astrophysical Sciences, Peyton Hall, Princeton University, Princeton, NJ 08544, USA}
\affiliation{ NASA Hubble Fellowship Program, Einstein Fellow}

\author[0000-0002-4584-2557]{Oliver Porth}
\affiliation{Anton Pannekoek Institute for Astronomy, University of Amsterdam, Science Park 904, 1098 XH, The Netherlands}

\author[0000-0002-1009-2354]{Anna Watts}
\affiliation{Anton Pannekoek Institute for Astronomy, University of Amsterdam, Science Park 904, 1098 XH, The Netherlands}



\begin{abstract}
The pulsed X-ray emission from the neutron star surface acts as a window to study the state of matter in the neutron star interior. For accreting millisecond pulsars, the surface X-ray emission is generated from the `hotspots' formed due to the magnetically channeled accretion flow hitting the stellar surface. The emission from these hotspots is modulated by stellar rotation giving rise to pulsations. Using global three-dimensional general relativistic magnetohydrodynamic (GRMHD) simulations of the star-disk system, we investigate the accretion hotspots and the corresponding X-ray pulse properties of accreting millisecond pulsars with dipolar magnetic fields. The accretion spot morphologies in our simulations are entirely determined by the accretion columns and vary as a function of the stellar magnetic inclination. For lower magnetic inclinations, the hotspots are shaped like crescents around the magnetic axis and are transformed into elongated bars for higher inclinations. We model the X-ray pulses resulting from the simulated hotspots using general-relativistic ray tracing calculations and quantify the variability of the pulsed signal. The pulse amplitudes in our simulations usually range between $1 - 12 \%$ rms and are consistent with the observed values. We find that the turbulent accretion flow in the GRMHD simulations introduces significant broadband variability on a timescale similar to the stellar rotational period. We also explore the impact of electron scattering absorption and show that along with being a key factor in determining the pulse characteristics, this also introduces significant additional variability and higher harmonics in the bolometric light curve of the accreting sources.
\end{abstract}



\section{Introduction} \label{sec:intro}
Accreting millisecond X-ray pulsars (AMXPs) are neutron stars in low-mass X-ray binaries (LMXBs) with relatively weak magnetic fields ($10^8 - 10^9$ G) spinning at frequencies larger than 100 Hz.  The surface X-ray emission mainly comes from the ``hotspots'', which are formed as a result of a magnetically channeled accretion flow. As the star rotates, these hotspots then give rise to X-ray pulsations (see \citealt{PatrunoandWatts2012} for a review). The surface emission of the AMXPs provides a window to study plasma flow in strong gravity, as well as to understand the dense matter equation of state (EOS). 
  
The properties of the resulting pulse profile (the rotational phase-resolved count spectrum) are set in part by relativistic effects that depend on the neutron star's mass and radius.  Pulse profile modeling \citep[PPM, ][]{Watts19}, a technique being applied to rotation-powered millisecond X-ray pulsars (RMPs) using data from NASA's Neutron Star Interior Composition Explorer \citep[NICER, ][]{Gendreau16}, takes advantage of these relativistic effects to enable us to infer not only the neutron star's mass and radius but also the geometric properties of the hotspots \citep[see for example][]{Vinciguerra24,Choudhury2024,Salmi2024,Dittmann2024}.  Mass and radius can be used to constrain the EOS \citep{Lattimer21}, while the hotspot properties provide insight into the magnetic field geometry of the star \citep[see e.g.][]{Bilous2019,Kalapotharakos21}. 

PPM can also be applied to AMXPs \citep{Salmi18,Dorsman2025}, which is an important science goal for future large-area X-ray spectral timing telescopes \citep{Watts2016}, but this brings new challenges. The surface emission mechanism and hotspot properties are different to the RMPs, being formed by the channeled accretion flow rather than return current heating. This indicates that the surface pattern models used in RMP PPM need to be modified for the AMXPs.  And unlike the RMPs, AMXP pulse profiles are variable, with pulse fractional amplitudes, harmonic content, and pulse phases varying significantly over time \citep[see e.g.][]{Hartman2008, Patruno2009, Patruno2010}.  

Several models have been proposed to explain these pulse variations. Magnetospheric coupling has been suggested to explain the pulse shape and phase variations in some of the sources \citep{Kajava2011}. Inner disk expansion \citep{Ibragimov2009} and hotspot movement \citep{Lamb2009} have been proposed to explain the phase drifts in some of the sources. Recently, scattering in the accretion column has also been suggested to have a significant impact on the pulse evolution \citep{Ahlberg2024}. However, all of the existing models are subject to strong assumptions on the geometry or nature of star-disk magnetospheric coupling, and hence, magneto-hydrodynamic (MHD) simulations are needed to check and complement the existing models. Global MHD simulations of turbulent accretion onto the neutron stars are particularly useful since they can provide a full global dynamical model of the emission geometry.

Non-relativistic MHD simulations of accreting stars with $\alpha$-viscosity  have been previously employed to study the accretion hotspots and the corresponding pulse properties \citep{Romanova2004, Kulkarni2008, Bachetti2010, Kulkarni2013}.  Although magnetorotational instability (MRI) driven star-disk simulations have been performed to investigate the accretion dynamics for smaller ($< 30^{\circ}$) stellar magnetic inclinations \citep{Romanova2011, Romanova2012}, the pulse properties resulting from the magnetized disk accretion have not yet been studied. General-relativistic magnetohydrodynamic (GRMHD) simulations of accretion onto stars with inclined magnetospheres have been only recently performed by \cite{Das2024, Murguia-Berthier2024} for dipolar stellar magnetic fields. \cite{Das2024} mainly focused on neutron star jets; in this paper, we investigate the processes near the stellar surface and discuss the properties of the accretion hotspots and the pulsed X-ray emission from the surface.   

The paper is structured as follows: Section \ref{sec:setup} describes the numerical setup for the GRMHD simulations and the ray-tracing calculations. We investigate the hotspot properties resulting from our GRMHD simulations in Section \ref{sec:hotspot-shapes}. We then perform ray-tracing calculations to study the pulse properties for the static averaged spot shapes in Section \ref{sec:avg_pulseprofiles} and explore pulse variability in Section \ref{sec:variability}. Finally, we study the impact of the accretion column on the pulse profiles, taking into account electron scattering absorption in Section \ref{sec:columns}. We discuss the results and conclude in Section \ref{sec:discussion}.

\section{Numerical Methods} \label{sec:setup}
In this section, we present the numerical setup for GRMHD simulations and the codes employed to generate the pulse profiles.
\subsection{GRMHD Simulations}
We solve the following GRMHD equations using the Black Hole Accretion Code ($\bhac$) \citep{Porth2017, Olivares2019} in Cartesian Schwarzschild coordinates corotating at the stellar angular frequency ($\Omega_{\rm star}$),
\begin{align}
    \nabla_{\mu}(\rho u^{\mu}) &= 0 \\
    \nabla_{\mu} T^{\mu\nu} &= 0 \\
    \nabla_{\mu} ^{\star}F^{\mu\nu} &= 0 \, .
\end{align}
Here, $\rho$, $u^{\mu}$, are rest-mass density and fluid four-velocity; $T^{\mu\nu}$ is the energy-momentum tensor of an ideal magneto-fluid and $^{\star}F^{\mu\nu}$ is the dual of the Faraday tensor $F^{\mu\nu}$. 

We initialize the simulation domain with a standard Fishbone - Moncrief torus \citep{Fishbone1976}, with the inner edge and density maximum located at $r_{\rm in} = 45 \, r_g$ ($r_g = GM/c^2$) and $r_{\rm max} = 65 \, r_g$ respectively. Inside the torus, we introduce poloidal magnetic field loops defined as $\rm A_{\phi} \propto max(\rho/\rho_{\rm max} - 0.1, 0)$ such that $2 \, p_{ \rm max}/b^2_{ \rm  max} \approx 110$, and outside the torus, the domain is initialized with a magnetic dipole centered at the coordinate center (0,0,0) following \citet{Wasserman1983}. Here $p$ and $b^2$ represent the pressure and the square of the magnetic field in the fluid frame respectively. In the magnetosphere ($r < r_{\mathrm{lc}} =$ $ c/\Omega_{\rm star}$), the initial pressure and density are set such that magnetization ($\sigma_t = b^2/\rho$) and plasma beta ($\beta_t = 2 \, p/b^2$) are 70 and 0.014 respectively. Both $\sigma$ and $\beta$ transition smoothly to follow an $r^{-6}$ profile outside the light-cylinder radius. We perturb the pressure with 8$\%$ white noise to excite the MRI inside the torus. The development of MRI results in angular momentum transport in the disk \citep{Balbus1991} and leads to accretion onto the neutron star. The simulations analyzed in this paper are described in more detail in \cite{Das2024}.

\par
Unless mentioned otherwise, we always use geometric units with $\textrm G = c = 1 $. The stellar radius, magnetic field strength (denoted by the dipole moment $\mu$), and rotational frequency for all the GRMHD runs are set to $\rstar = 4 \, r_g$, $\mu = 20 \, \mu_{cgs}$, and $\Omega_{\rm star} = 0.03\, c/r_{g}$ respectively, where $\mu_{cgs}$ is $\sqrt{ 4\pi \rho_{\rm cgs}c^2}\, r_g^3$. Finally, assuming $\dot{M} = 1$ [code units] corresponding to $1\% \dot{M}_{\rm Edd}$ results in a physical density scaling of $\rho_{\rm cgs} = 1.27 \times 10^{-5}$ g cm$^{-3}$ for a radiative efficiency of 0.1. Here, $\dot{M}_{\rm Edd}$ denotes the mass accretion rate corresponding to the Eddington luminosity.
We extract all the relevant quantities on a spherical shell with radius $r = 4.3 \, r_g$. This ensures that the interpolation stencil stays clear of the stellar interior where the solution is excised. Thus, the equatorial radius of the neutron star ($R_{\rm eq}$) is always assumed to be $4.3 \, r_g$ for the ray-tracing calculations. In order to extract observable signatures from the GRMHD simulations, we scale our simulations to physical units following \cite{Das2022}. Assuming $\rstar = 10$ km, allow us to convert the stellar parameters to the following physical units, $M = 1.69 \, {M}_{\odot}$, $R_{\rm eq}$ = 10.75 km, which sets $\Omega_{\mathrm{star}} = 572.56$ Hz. The pulses are computed for a range of observer frequencies, $\nu_{\rm {obs}} \in [2.42 \times 10^{15}, 1.23 \times 10^{19}]$ Hz ([0.01, 50.87] KeV). 
\subsection{Raytracing}
To compute the pulses from the stellar surface, we perform general relativistic ray-tracing calculations using $\raptor$ \citep{Bronzwaer2018, Bronzwaer2020} and  $\xpsi$ \citep{xpsi} in a static Schwarzschild spacetime. We initially perform the ray-tracing calculations for the time-averaged surface profiles using $\xpsi$ and finally use $\raptor$ to fully explore the impact of time-dependent accretion dynamics on the surface emission. Both these codes use different ray-tracing approaches which are further explained below.

\subsubsection{RAPTOR}
$\raptor$ solves the following geodesic equations numerically for a given initial position (${x^{\alpha}}_0$), and initial contravariant wave vector ($k^{\alpha}_0$),  
\begin{align}
& \frac{\mathrm{d} x^\alpha}{\mathrm{d} \lambda}=k^\alpha, \\
& \frac{\mathrm{d} k^\alpha}{\mathrm{d} \lambda}=-\Gamma_{\mu \nu}^\alpha k^\mu k^\nu \, .
\end{align}
Here, $\lambda$ is the affine parameter; $x^{\alpha}$, $k^{\alpha}$, and $\Gamma_{\mu v}^\alpha$ are the position, contravariant wave vector, and Christoffel symbols, respectively. The initial wave vectors are constructed following \cite{Cunningham1972}. The initial positions of the photon are always set by the location of the virtual camera which is set to ($r, \theta, \phi$) = $(10^4 \, r_g, \iobs, 0)$, where $\iobs$ represents the observer inclination. The numerical integration is performed using a 4th-order Runge-Kutta method. In this ray-tracing method, the photons are traced backward from the virtual camera, and thus, also known as the \textit{backward integration} method. In \raptor, we assume a fast-light approximation where the geodesic’s time coordinate is ignored. The impact of ignoring time delays on the pulse profiles is shown for the test cases in Section \ref{sec:codetests}. The base resolution of the virtual camera is set to $400 \times 400$ pixels with 2 levels of refinement resolving the hotspots with an effective of $800 \times 800 $ pixels\footnote{We have performed a convergence study which showed that the difference in spectrum stays within $0.25 \%$ for the runs with $1600 \times 1600$ pixels and $400 \times 400$ pixels with 2 refinement levels.} (See \citealt{Davelaar2022} for a detailed description of the camera refinement implementation in \raptor). The camera in \raptor is a block-based quadtree, for each block, we first compute the emission at the lowest refinement level. Then by taking the gradient of the specific intensity sharp features are identified. If a sharp feature is present, the block will split into four, with two new blocks in each direction, each containing the same number of pixels. This is done iteratively until the sharp features are either fully resolved or when a user-defined maximum depth is achieved. This way is reminiscent of the adaptive mesh refinement strategies exploited in the GRMHD codes.\par
Regarding the emission at the stellar surface, we compute the fluid frame (comoving with the plasma) specific intensity ($I_{\nu}$) from the GRMHD simulations and get the observer intensity following, 
\begin{align}\label{eq:obs_intensity}
    I_{\rm \nu, obs} = \frac{I_{\nu}}{\nu^3} \nu_{\rm obs}^3 \, .
\end{align}
Here $\nu$ is the fluid frame frequency given by $\nu = - k^{\alpha} u_{\alpha}$ and $u_{\alpha}$ is the fluid four-velocity. At the stellar surface, the four-velocity is  given by $u^i = u^{t}[1,0,0,\Omega_{\rm star}]$, where $u^{t}$ is obtained from $u^i u_i = -1$.\par
Throughout this paper, we set $I_{\nu}$ to be the Planck function, i.e., we assume isotropic black-body radiation.  
In reality, the observed AMXP spectra also exhibit a power-law component likely produced by the Comptonization of thermal seed photons in a region close to the hotspot \citep[see e.g.,][]{Poutanen2003,Bobrikova2023}.
This component has an anisotropic beaming pattern affecting also the shape of the pulses. The Comptonization component likely originates from the radiation-dominated shock above the stellar surface \citep{BaskoSunyaev1976}. Since our stellar boundary conditions are set such that the accreting materials flow through the surface unhindered, we do not model the shock and hence, also ignore the corresponding Comptonization component. 
\par
For neutron stars rotating at considerably high frequencies ($\Omega_{\rm star} > 200$ Hz), the oblateness of the surface is significant and begins to impact the corresponding pulse profiles. To incorporate the oblate stellar surface in $\raptor$, we modify the initially spherical stellar surface to an oblate spheroid following \cite{AlGendy2014} such that, 
\begin{equation}
R\left(\theta\right)=R_{\mathrm{eq}}\left[1+o_2(x, \bar{\Omega}) \cos ^2\left(\theta\right)\right] \, .
\label{eq:oblateness}
\end{equation}
Here, $R_{\rm eq}$ is the equatorial radius and $o_2(x, \bar{\Omega}) = \bar{\Omega}^2\left(-0.788 + 1.030 x\right)$ with $x = M/R_{\rm eq}$, $\bar{\Omega} = \Omega_{\rm star} (R_{\rm eq}^3/M)^{1/2}$. \par
Finally, we solve the following radiative transfer equations for the Lorentz invariant quantities \citep{Lindquist1966} along the null geodesics to understand the impact of absorption in the accretion columns on the surface emissions. 
\begin{align}
\frac{\mathrm{d}}{\mathrm{d} \lambda}\left(\frac{I_{\nu}}{\nu^3}\right)=-\nu \alpha_{\nu}\left(\frac{I_{\nu}}{\nu^3}\right) \, .
\label{eq:RT}
\end{align}
Here $I_{\nu}$ is the specific intensity and $\alpha_{\nu}$ is the absorption coefficient. To explore the effects of electron scattering in the accretion column on the pulse profiles, we set the absorption coefficient to Thomson scattering opacity for pure hydrogen ($ \alpha_{\nu} = 0.4$ $\rm cm^2$ $\rm g^{-1}$).  
\subsubsection{X-PSI}
In \xpsi, we use a \textit{forward integration} method, where photons are traced from the NS surface to the observer in a static Schwarzshild space-time with the special relativistic corrections applied to the observed fluxes and the oblateness of the star taken into account following the Oblate Schwarzschild (OS) approximation \citep[for details of the OS approximation see][]{Morsink2007, Bogdanov2019_PPM}.
The ray-tracing is performed by solving the light bending angles $\psi$ from the star's surface to the observer at infinity for a set of NS radii $R(\theta)$ and emission angles $\alpha$ (measured between the radial direction and initial direction of the emitted photon), given by the following relation \citep[when $\alpha < \pi/2$,][]{mtw73,Pechenick1983}:
\begin{equation}\label{eq:light_bending}
  \psi_{\mathrm{p}}(R,\alpha)=\int_R^{\infty} \frac{dr}{r^2} \left[ \frac{1}{b^2} -
       \frac{1}{r^2}\left( 1- \frac{2r_g}{r}\right)\right]^{-1/2} ,
\end{equation}
where 
\begin{equation} \label{eq:impact_parameter}
  b=\frac{R}{\sqrt{1-2r_g/R}} \sin\alpha
\end{equation}
is the impact parameter. 
For a photon to reach the observer, the angle $\psi$ is already known given the viewing geometry and hotspot location. 
The emission angle $\alpha$ corresponding to that $\psi$ is then interpolated from a precomputed set of rays.
These angles are then used to calculate the observed flux, after accounting for special relativistic corrections and the oblate shape of the star (see Equation \eqref{eq:oblateness}).

The observed flux is obtained from 
\begin{equation}\label{eq:fluxs_xpsi}
dF_{\nu,\mathrm{obs}}=(1-2r_g/R)^{1/2} \delta^{3} I'_{\nu'} \cos\sigma'
\abs[\Big]{\frac{d \cos\alpha}{d \cos\psi}}_{R}
 \frac{dS'}{D^2} ,
\end{equation}
\citep{Bogdanov2019_PPM} where the primed quantities are measured in the local surface comoving frame (so at the stellar surface $I'_{\nu'}$ is equivalent to $I_{\nu}$ mentioned in Equation \eqref{eq:obs_intensity}), $D$ is the distance to the star, $\sigma'$ is the emission angle relative to the surface normal, $dS'$ is a surface area element, and $\delta$ is the Doppler factor that depends on the NS surface velocity and the angle between the velocity and the direction of the emitted photon.
The effect of time delays due to different photon paths is usually accounted for in the \xpsi calculation (unlike in \raptor), but in some cases in Section \ref{sec:codetests} it is turned off to match \raptor results.
\footnote{Ignoring the geodesic's time coordinate in \raptor affects also the observed number of photons per second even for a single photon path.
Thus, when ignoring time delays, we also neglect one of the Doppler factors in Equation \eqref{eq:fluxs_xpsi}, coming from the photon arrival time contraction \citep[see e.g.,][]{Rybicki1979,Poutanen2006}.}
The calculations were based on \xpsi version \texttt{v1.2.1}\footnote{\url{https://github.com/xpsi-group/xpsi}} modified additionally to allow surface temperature maps from the GRMHD simulations.


\subsubsection{Code Tests}\label{sec:codetests}
\begin{figure*}
    \centering
    \text{Spherical Star} \\
    \begin{tabular}{c c}
        \includegraphics[width=8cm, angle=0]{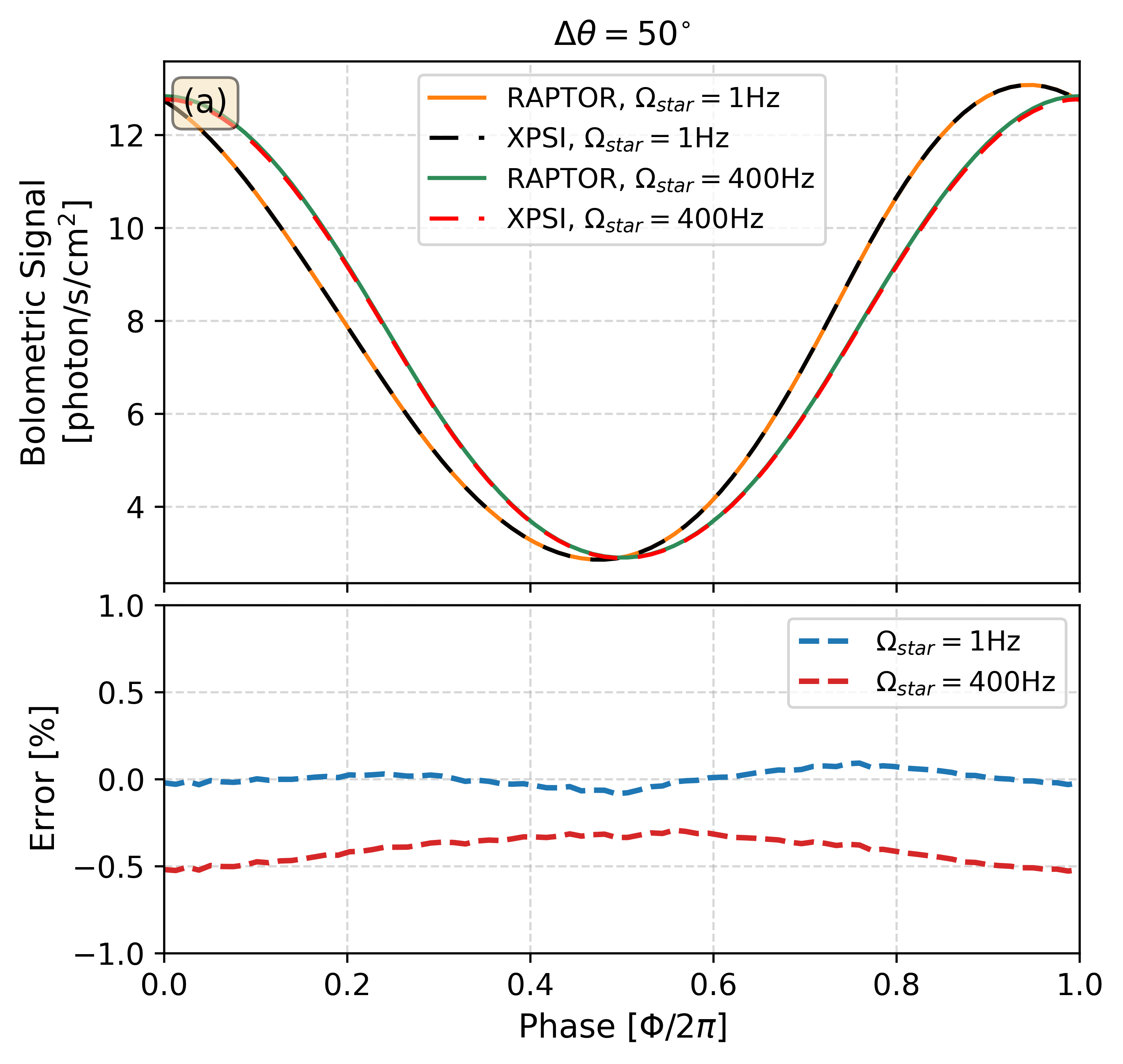} &
        \includegraphics[width=8cm, angle=0]{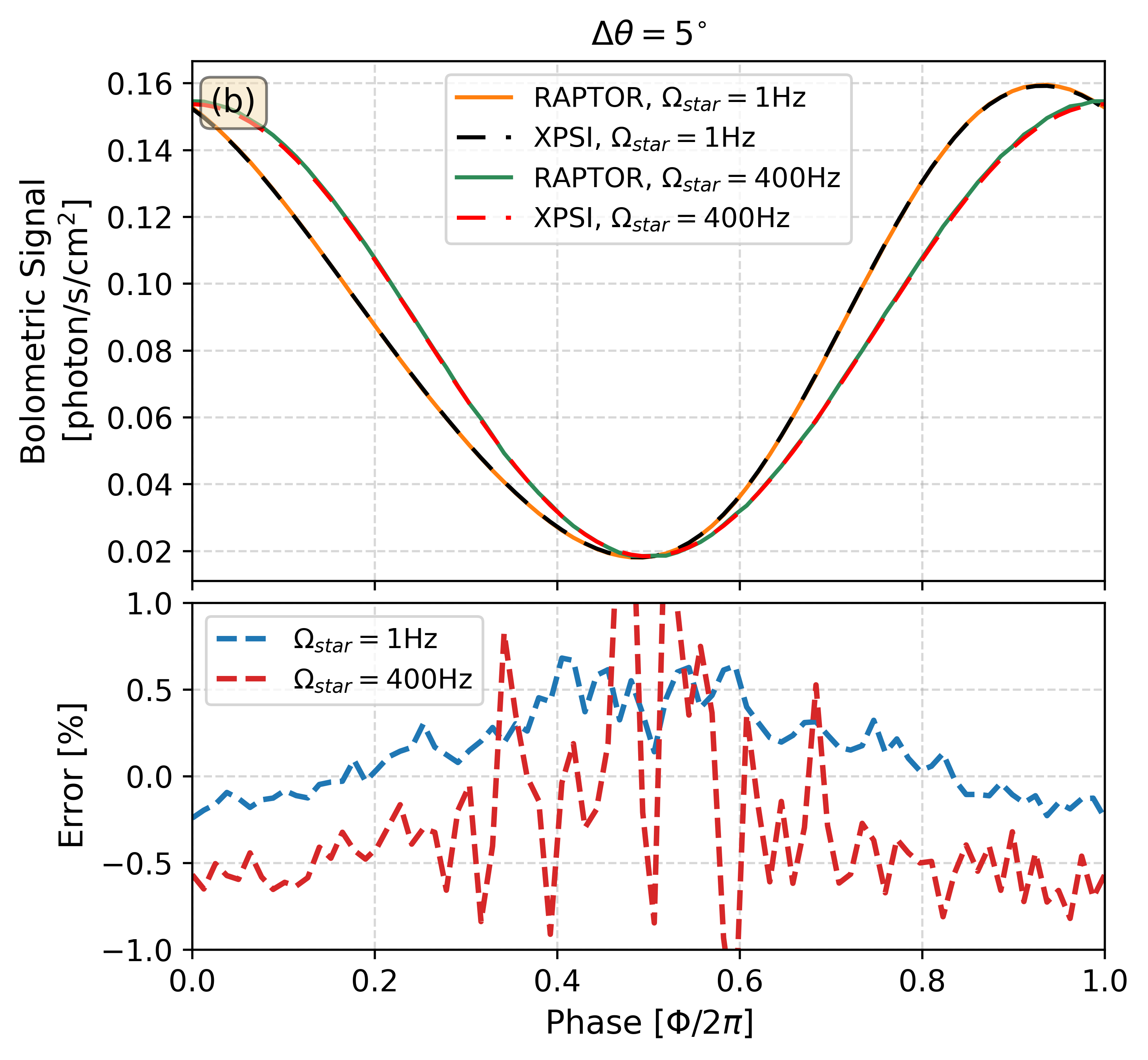} \\
    \end{tabular}\\
    \text{$\Omega_{\mathrm{star}} = 400$ Hz, Oblate Star} \\
    \begin{tabular}{c c}
        \includegraphics[width=8cm, angle=0]{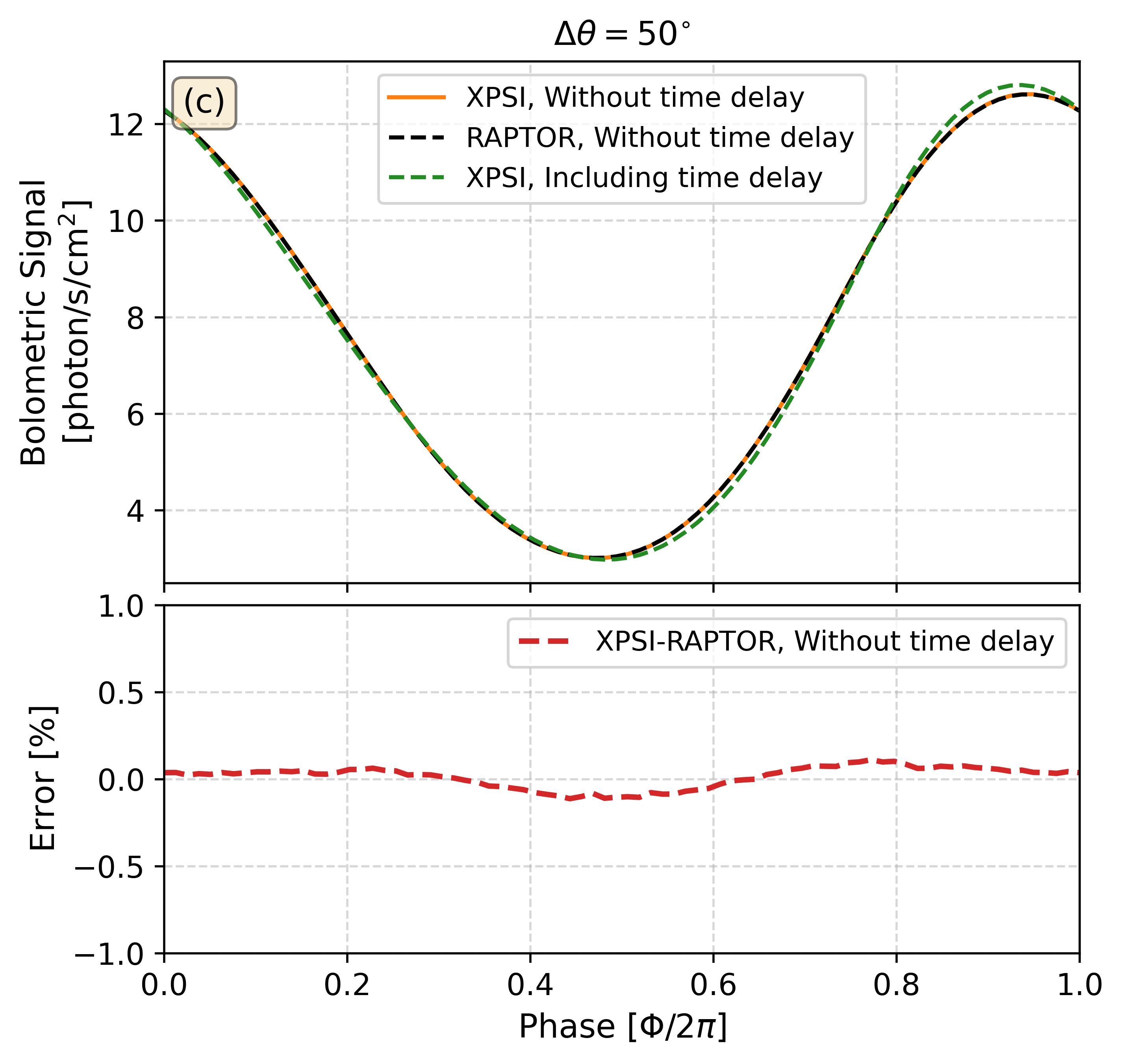} &
        \includegraphics[width=8cm, angle=0]{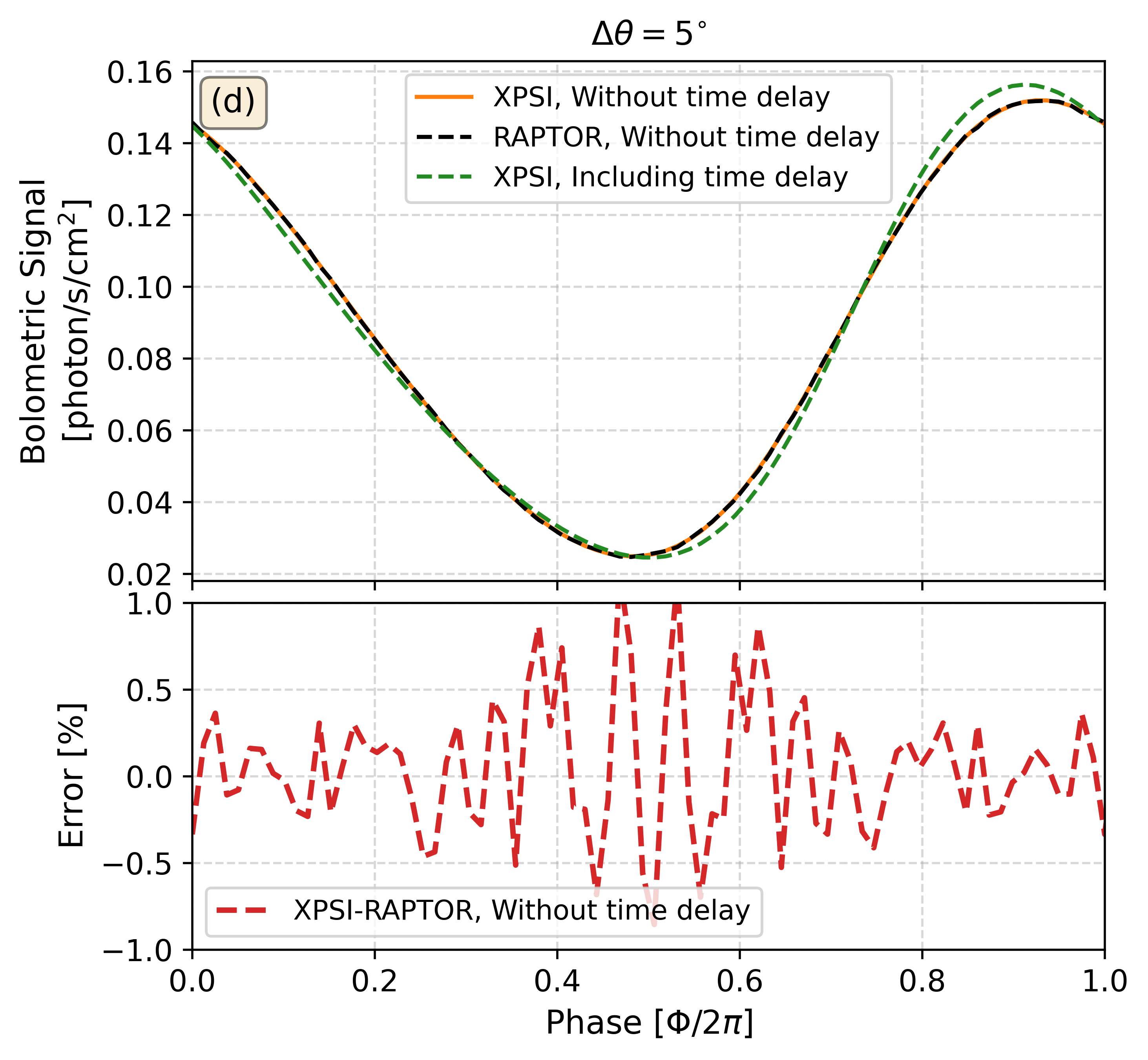} \\
    \end{tabular}
    \caption{Bolometric flux for $\raptor$ and $\xpsi$ with a uniform temperature spot ($kT_{\rm eff} = 0.35$ keV). The center of the hotspot for all the tests is at $\theta_c = 50^{\circ}$. The two columns show the bolometric pulse profiles for two different spot sizes ($\Delta \theta$), $50^{\circ}$ and $5^{\circ}$ respectively. The first row (a, b) shows the tests for a spherical star rotating at 1 Hz and 400 Hz. The second row (c,d) shows the bolometric profiles for an oblate star rotating at 400 Hz. The observer inclination for all the cases is set to $i_{\mathrm{obs}} = 60^{\circ}$. The lower panels in each of the plots show the fractional errors between $\raptor$ and $\xpsi$. The green dashed line in the lower panel shows the bolometric profile with time delays.}%
    \label{fig:codetests}
\end{figure*}

To check the agreement between the two ray-tracing codes, we compared the pulse profiles obtained with $\raptor$ and $\xpsi$ for a few test cases. We have considered black-body emission ($kT_{\rm eff} = 0.35$ keV) from a single spot at the stellar surface for a star with $M = 1.4 ~{M_{\odot}}$ and $R_{\mathrm{eq}} = 12$ km at $D = 0.2$ kpc for two different spot sizes, $\Delta \theta = 5^{\circ}$ and $50^{\circ}$ with the spot centers fixed at $\theta_c = 50^{\circ}$. The observer inclinations for all the test cases are set to \iobs $= 60^{\circ}$. The comparisons for both slow ($\Omega_{\mathrm{star}} = 1$ Hz) and fast ($\Omega_{\mathrm{star}} = 400$ Hz) rotating spherical stars are shown in \autoref{fig:codetests}(a,b). The second row in \autoref{fig:codetests}(c,d) shows the profiles for an oblate star rotating at $\Omega_{\mathrm{star}} = 400$ Hz, where the oblateness of the surface is incorporated following Equation \eqref{eq:oblateness}. We observe that in all cases, the difference in the bolometric flux between $\raptor$ and $\xpsi$ stays $\lesssim 1\%$ in almost all the phases, demonstrating an excellent agreement between the two codes. 
The green curve in \autoref{fig:codetests}(c, d) shows the impact of including time delay on the pulse profiles. The difference in the pulse profiles with and without time delays is approximately $\leq 5\%$.  
\section{Results} \label{sec:Result}
\subsection{GRMHD Simulations}\label{sec:hotspot-shapes}
For our chosen magnetic field strength, in the case of the aligned stellar magnetic field ($\chi_{\mathrm{star}} = 0^{\circ}$), the disk accretes through two symmetric accretion columns in the upper and lower magnetic hemispheres \citep[see Figure 1 in ][]{Das2024}. However, with increasing magnetic inclination, the accretion columns exhibit a growing asymmetry around the magnetic axis, and most of the disk material gets accreted through the column close to the equatorial disk in each magnetic hemisphere (\autoref{volume_rendering}). The accreted material is deposited onto the stellar surface, resulting in the formation of hotspots. A detailed description of the magnetospheric dynamics which determines the accretion column formation is provided in \cite{Das2024}. \par
Assuming that the entire matter-kinetic energy flux ($F_{\rm m}$) of the accretion column is emitted as black-body radiation at the stellar surface leads to 
\begin{align}
    T_{\rm eff} &= \bigg(\frac{F_{\rm m}}{\sigma_{\rm 
 SB}}\bigg)^{1/4},
    \label{eq:temp}
    \end{align}
    where
    \begin{align}
    F_{\rm m} &= {T^{\text{MA}}}^r_t = (\rho_0 + u_g + p_g)u^r u_t \, . 
    \label{eq:flux}
\end{align}
Here, $F_{\rm m}$ is the total matter-energy flux; ${T^{\text{MA}}}^r_t$ is the matter component of the stress-energy tensor \citep{McKinney2012}; $T_{\rm eff}$ and $\sigma_{\rm SB}$ are the effective black-body temperature and Stefan-Boltzmann constant respectively. Finally, assuming $\dot{M}=1$ [code units] corresponds to $1\%\dot{M}_{\rm Edd}$ (radiative efficiency = 0.1), we compute the temperature in physical (cgs) units. \par
\begin{figure*}
    \centering
    \includegraphics[width=16cm]{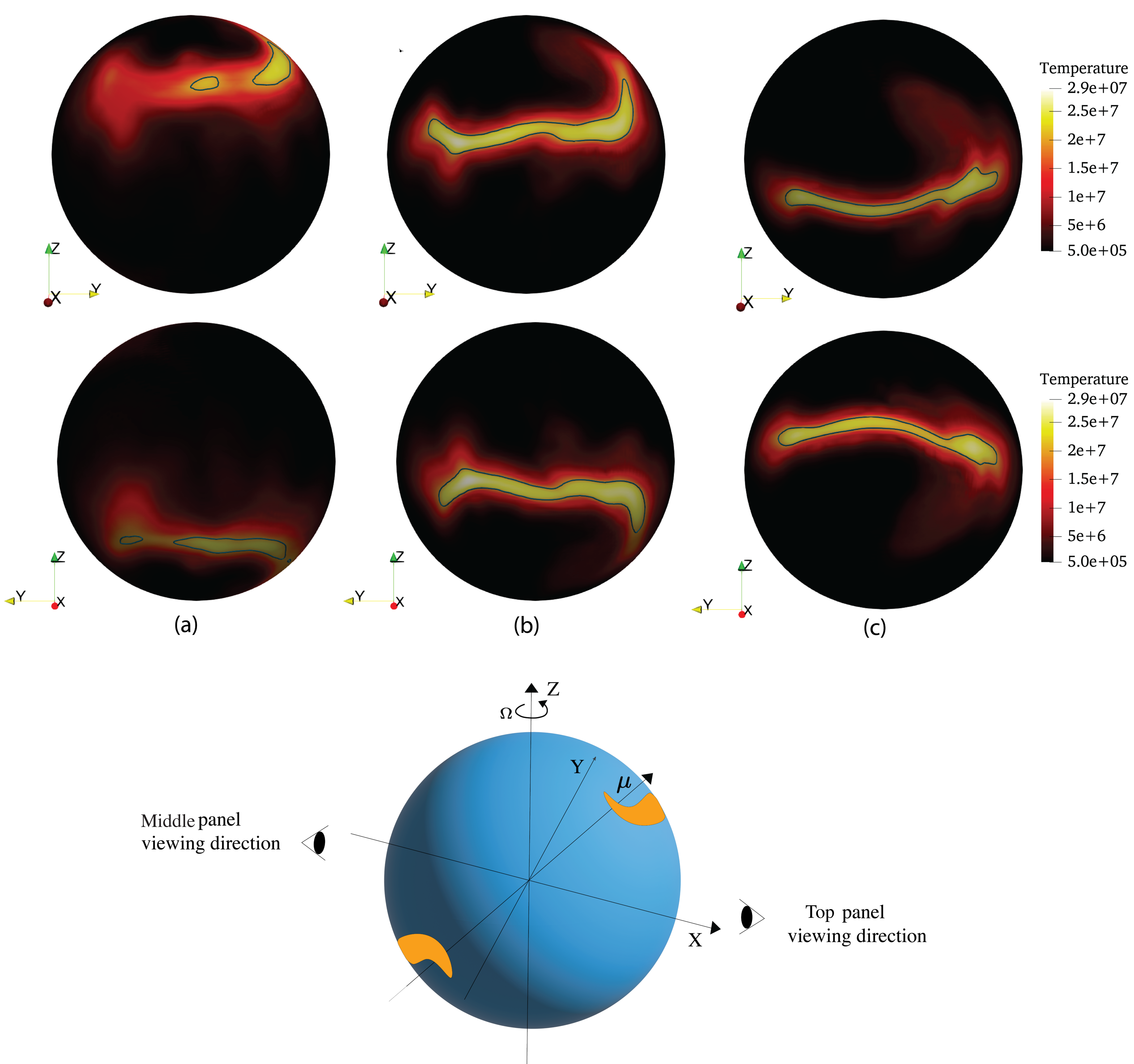}%
    \caption{Averaged hotspot shapes at the stellar surface ($r = 4.3 \, r_g$) for (a) $\chi_{\mathrm{star}} = 30^{\circ}$, (b) $\chi_{\mathrm{star}} = 60^{\circ}$ and (c) $\chi_{\mathrm{star}} = 90^{\circ}$ with the magnetic axis in the $z-x$ plane. The top and the middle panels show the spot structures in the upper and lower magnetic hemispheres respectively. All the shells are averaged over t $\in [15000,25000] \, r_g/c \approx 48$ stellar periods. The color bar shows temperature in Kelvin (converted to physical units assuming $\dot{M}$ scaling as 1$\%$ Eddington accretion rate). The blue contours outline the regions with temperatures $> 2.0 \times 10^7$K (1.72 keV). The cartoon (blue sphere) depicts the viewing direction of the hotspots in the top and middle panels.}%
    \label{fig:surfacemaps_avg}
\end{figure*}
\autoref{fig:surfacemaps_avg} shows the hotspot structures on a spherical surface at $4.3r_{\rm g}$, for different stellar magnetic inclinations ($\chi_{\mathrm{star}} \in [30^{\circ}, 60^{\circ}, 90^{\circ}]$). The flux is averaged over t $\in [15 000, 25 000] \, r_g/c$.  This averaging window roughly corresponds to a viscous timescale at $r=8 r_{\rm g}$ -- the effective magnetospheric radius of the simulations \citep{Das2024}.
The location and shape of the hotspots in our simulations are entirely determined by the accretion columns. As a visual aid, we show volume renderings of the column structure for  $\chi_{\rm star} = 60^{\circ}$ and $\chi_{\rm star} = 90^{\circ}$ in \autoref{volume_rendering}. As we start increasing the magnetic inclination ($\chi_{\rm star} \gtrsim 30^{\circ}$), the non-axisymmetric star-disk magnetic field interaction leads to a variation in the column structures along the azimuthal direction. In the $\Omega - \mu$ (i.e., in the $z-x$) plane, the gas is mainly accreted via one dominant accretion column along the magnetic pole closest to the equatorial disk. However, in the regions close to the $\mu - y$ plane, the equatorial disk forms two symmetric accretion columns (with respect to the magnetic equator) and leads to accretion along both the magnetic poles. This asymmetry in the column structures is one of the main reasons behind the non-uniform hotspot widths along the azimuthal plane.

For $\chi_{\mathrm{star}} = 30^{\circ}$, the spots are mainly in the shapes of crescents near the magnetic axis and transform into elongated bands with increasing $\chi_{\mathrm{star}}$. Finally for $\chi_{\mathrm{star}} = 90^{\circ}$, the spot shape resembles a narrow belt close to the equator. 
For the lowest magnetic inclination ($\chi_{\mathrm{star}} = 30^{\circ}$), our hotspot shapes closely resemble those found in \cite{Romanova2012}, and the variations between the two can be attributed to differences in the accretion flow in the inner magnetosphere. 
\begin{figure*}
    \centering
    \includegraphics[width=18cm]{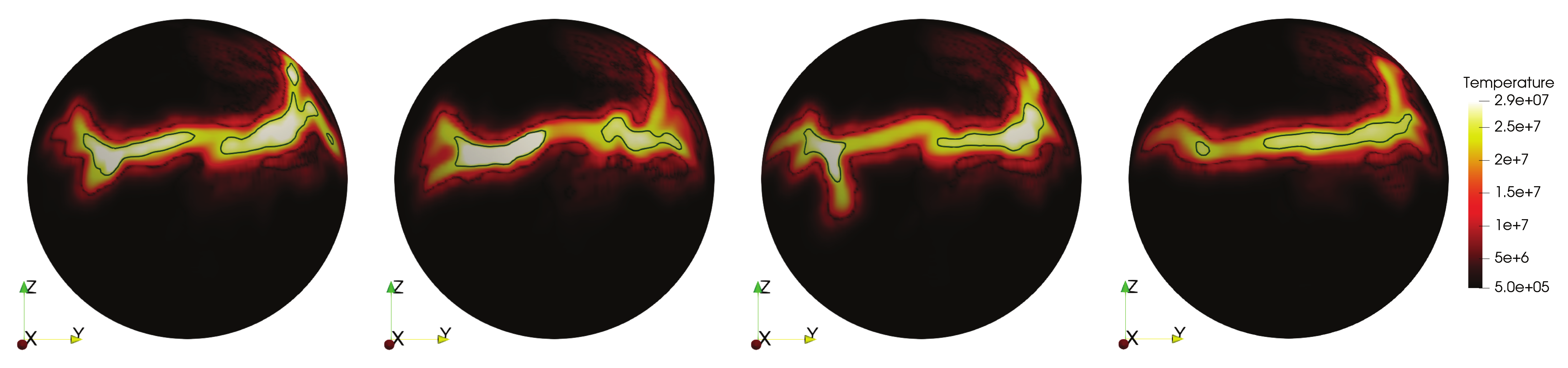}%
    \caption{Variability of temperature profiles and spot shapes for $\chi_{\mathrm{star}} = 60^{\circ}$ around $1200 r_{\textrm{g}}/c$ ($\approx 10$ ms) apart. The variation in the high-temperature regions ($T_{\mathrm{eff}} > 2.5 \times 10^7$ K) is outlined with the green contours. An animated version of the figure, illustrating the temperature evolution over $\rm 83 \, ms$ with a cadence of $\rm 0.17 \, ms$, is available at: \url{https://youtu.be/7QD3z1SKESE}.}%
    \label{fig:surfacemaps_10ms}
\end{figure*}

The MRI-driven accretion variability significantly influences both shapes and temperature gradients within the accretion column and, hence, the hotspots. \autoref{fig:surfacemaps_10ms} shows the spot shapes for $\chi_{\mathrm{star}} = 60^{\circ}$ every $1200 \, r_g/c$ ($\approx 10$ ms). The green contours in each panel highlight the movement of the high-temperature regions with $T_{\mathrm{eff}} > 2.5 \times 10^7$ K (2.15 keV)\footnote{This value is chosen arbitrarily to guide the reader's eye.} within the spot. We find that the turbulent accretion disk leads to variations in the spot shapes on a timescale similar to the stellar rotational period, especially at the edges where the changes in the star-disk connectivity of field lines are most pronounced. Since the disk forms almost symmetric columns in the $\mu - y$ plane, Rayleigh-Taylor instabilities are also more pronounced in those regions. 
The impact of these hotspot variations on the pulse profiles is further investigated in Section \ref{sec:variability}.
\subsection{Pulse Profiles}
In the following section, we use two different ray-tracing approaches to study the pulses obtained from the simulated accretion hotspots. First, we compute the pulse profiles from temporally averaged hotspot shapes using a forward-in-time method (light rays propagating from the star to the observer). We then ray-trace each GRMHD snapshot using a backward-in-time method (rays from observer to star) to take temporal variations of the accretion flow into account. 
\par
Although the shapes and the temperature profiles on the accreting hotspots vary on a similar timescale to the stellar rotation period (for example see \autoref{fig:surfacemaps_10ms}), as a first step, we study the pulses from the averaged hotspots for different observer inclinations (in Section \ref{sec:avg_pulseprofiles}). Next, we study the effect of time variability on the generated pulse profiles in Section \ref{sec:variability}.\par

\subsubsection{Averaged Pulse Profiles}\label{sec:avg_pulseprofiles}
To obtain the average temperature profile, we extract the matter-energy flux at the stellar surface and compute the temporal average for t $\in [15000, 25000] \, r_g/c$ ($\sim 47$ cycles). The temperature then follows from \autoref{eq:temp}. This average hotspot temperature is realized in the limit where the thermal timescale of the neutron star atmosphere is much longer than the timescale of accretion fluctuations. \par
\begin{figure}
    \centering
    \includegraphics[width=8.6cm, angle=0]{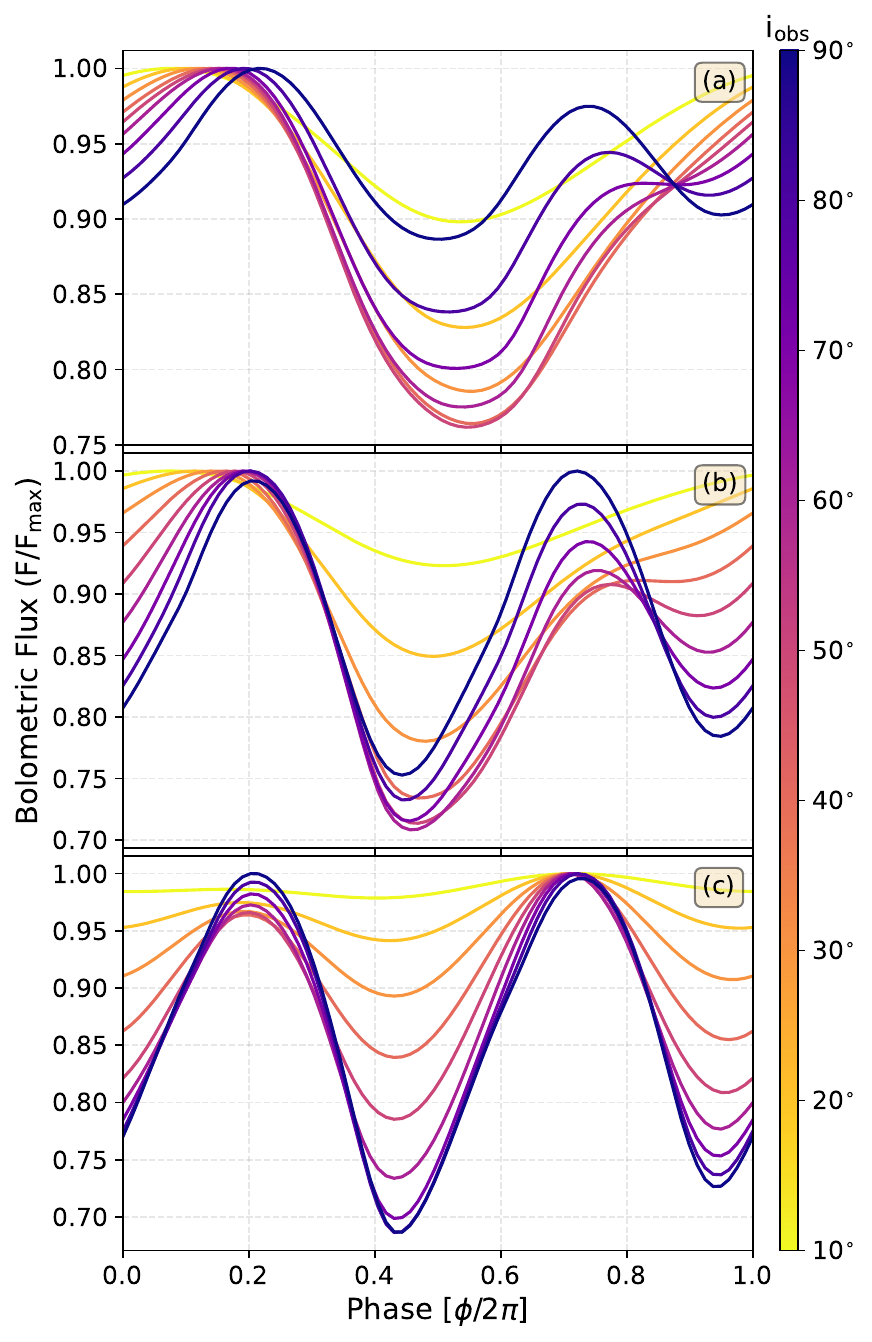}
    \caption{Normalized bolometric pulse profiles of averaged hotspot shapes (\autoref{fig:surfacemaps_avg}) for (a) $\chi_{\mathrm{star}} = 30^{\circ}$, (b) $\chi_{\mathrm{star}} = 60^{\circ}$ and (c) $\chi_{\mathrm{star}} = 90^{\circ}$. Here, each pulse is normalized to its maximum ($F_{\rm max}$). Different colors show the pulses for different observer inclinations.}%
    \label{fig:avgpp}
\end{figure}
The panels a, b, and c in \autoref{fig:avgpp} show the normalized pulses for three GRMHD simulations ($\chi_{\mathrm{star}} \in [30^{\circ}, 60^{\circ}, 90^{\circ}]$) obtained using \xpsi, with different colors representing pulses for different observer inclinations. We find that for all magnetic inclinations, the pulse fraction is lowest for \iobs $= 10^{\circ}$ (we vary \iobs $\in [10^{\circ}, 90^{\circ}]$), and increases with increasing observer inclination. For lower observer inclinations, i.e., for an observer close to the pole, the emission from the northern hotspot is always observable. Increasing $\chi_{\mathrm{star}}$ results in both spots moving closer to the equator, and as a result, the southern hotspot becomes increasingly visible irrespective of the phase.  This effect is considerable for  $\chi_{\mathrm{star}} = 60^{\circ}$, and finally, for $\chi_{\mathrm{star}} = 90^{\circ}$, both spots, being near the equator contribute almost equally and are always visible. Thus, for an observer close to the pole, we have a decreasing pulse fraction with $\chi_{\mathrm{star}}$.  As we increase \iobs,  (a) the pulse peak shifts in phase, and (b) an intermediate peak begins to appear. These two features are common across all the hotspot geometries. Increasing \iobs\ leads to only one spot being visible at a single stellar phase. With the antipodal hotspot geometry, the primary peak occurs due to the northern spot, and the intermediate pulse occurs as a result of the southern spot. 
\par 
To quantify the harmonic content of the pulses in \autoref{fig:avgpp}, we fit the following function, 
\begin{align}
    f(\phi) = A + B \sin{(\phi + \phi_1)} + C \sin{(2\phi + \phi_2)}.
    \label{eq:fit_harmonic}
\end{align}
where $B$ and $C$ are the fundamental and first harmonic amplitudes, and $\phi_1$ and $\phi_2$ are the corresponding phase shifts. Here, $A$ represents the average bolometric photon flux in the units of photons/s/cm$^2$. 
\par
\begin{figure}
    \centering
    \includegraphics[width=8.8cm, angle=0]{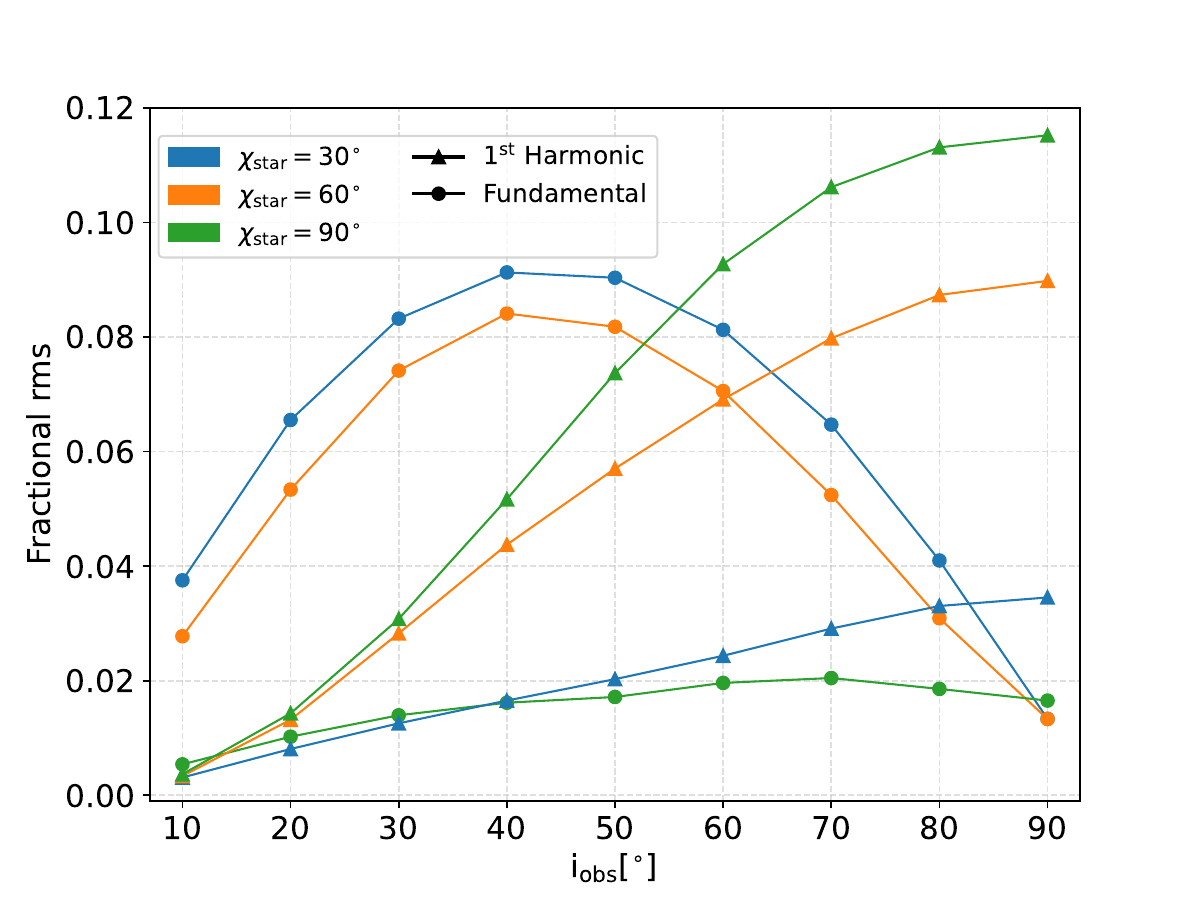}
    \caption{Fractional rms amplitude of the fundamental and the first harmonic of the averaged pulse profiles (\autoref{fig:avgpp}) as a function of observer inclination for $\chi_{\mathrm{star}} \in [30^{\circ}, 60^{\circ}, 90^{\circ}]$.}
    \label{fig:harmonic_content}
\end{figure}

The fractional rms amplitudes of the fundamental and first harmonic for all the cases are summarized in \autoref{fig:harmonic_content}. The residuals of the fit stay within $1.2\%$.  
For $\chi_{\mathrm{star}} = 30^{\circ}$ and $60^{\circ}$, the fractional rms amplitude of the fundamental (B/A$\sqrt{2}$) follows a similar trend with $i_{\mathrm{obs}}$. It increases from \iobs $= 10^{\circ}$ (0.037 and 0.027) to \iobs $= 40^{\circ}$ (0.091 and 0.084 for $\chi_{\mathrm{star}} = 30^{\circ}$ and $60^{\circ}$ respectively) and finally reduces back to $\sim 0.013$ as we further increase $i_{\mathrm{obs}}$ to $90^{\circ}$. For $\chi_{\mathrm{star}} = 90^{\circ}$, however, the rms amplitude of the fundamental steadily increases (from 0.005 to 0.016) with increasing $i_{\mathrm{obs}}$. The fractional rms of the first harmonic  (C/A$\sqrt{2}$) always increases with $i_{\mathrm{obs}}$ (from $\sim 0.003$ to 0.034, 0.089, and 0.115 for $\chi_{\mathrm{star}} = 30^{\circ}, 60^{\circ}$, and $90^{\circ}$, respectively). 
\subsubsection{Variability}\label{sec:variability}
Our simulations, in general, always give rise to highly variable mass accretion rates. At the stellar surface, this is manifested in terms of changing spot shapes and variable surface temperature profiles.
\begin{figure*}
    \centering
    \includegraphics[width=17.6cm, angle=0]{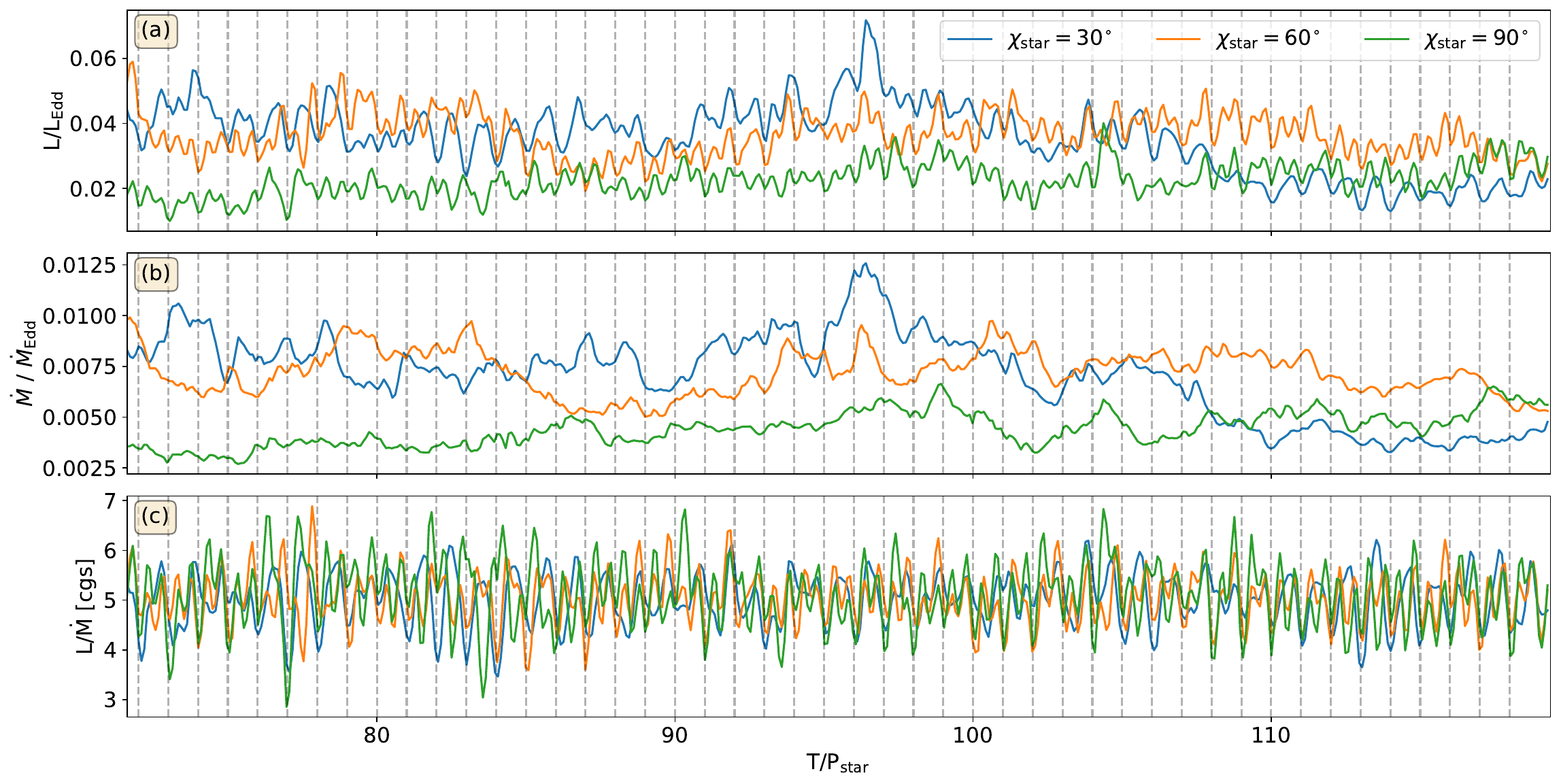}
    \caption{Time evolution of (a) bolometric luminosity ($L$), (b) mass accretion rate ($\dot{M}$), and (c) $\dot{M}$ normalized bolometric luminosity for different stellar magnetic inclinations. The observer inclination is set to $70^{\circ}$. The stellar periods are highlighted by the grey dashed lines.}%
    \label{fig:intensity-diffmag}
\end{figure*} 
The mass accretion rate is defined as follows,
\begin{align}
    \dot{M} &= \int_{0}^{2 \pi} \int_{0}^{\pi} \sqrt{-g} \rho u^r \,d\theta \,d\phi 
\end{align}
where $\rho$ and $u^{r}$ represent the density and the radial four velocity respectively. 
Panel (a) and (b) in \autoref{fig:intensity-diffmag} show the observed bolometric luminosity and the surface mass accretion rate for different magnetic inclinations for 47 stellar periods obtained using \raptor. In the given time period, the variance in the light curve ($L$) and the $\dot{M}$ time series can be measured by the coefficient of variation ($c_x$, or the total fractional rms variability) defined as, $c_x = \sigma_x/\langle x \rangle$ for a quantity $x$, with $\sigma_x$ and $\langle x \rangle$ representing the standard deviation and the mean respectively. For $\chi_{\mathrm{star}} \in [30^{\circ}, 60^{\circ}, 90^{\circ}]$, $c_{L} = [0.300, 0.182, 0.219]$ and $c_{\dot{M}} = [0.287, 0.151, 0.187]$ respectively. All the quantities are measured for t $\in [15000,25000] \, r_g/c$ and the observer inclination is always set to $70^{\circ}$. As expected, apart from additional rotational modulation, the time evolution of the observed luminosity closely follows the surface mass accretion rate. The bolometric luminosity always correlates positively with the mass accretion rate, with correlation coefficients 0.928, 0.820, and 0.804 for $\chi_{\mathrm{star}} \in [30^{\circ}, 60^{\circ}, 90^{\circ}]$ respectively.\par
\begin{figure}
    \centering
    \includegraphics[width=9.2cm, angle=0]{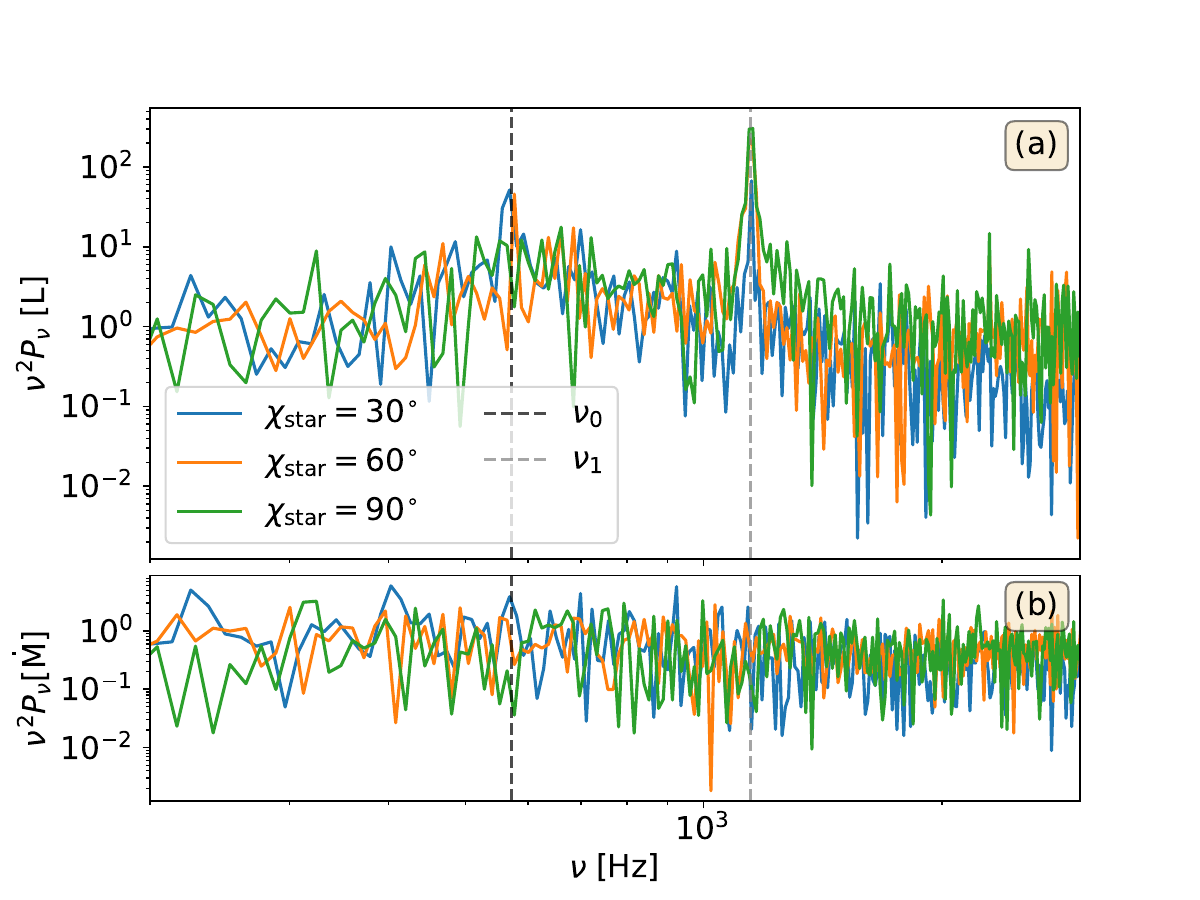}
    \caption{RMS-normalized power spectrum of (a) Luminosity ($L$) and (b) $\dot{M}$ for $\chi_{\mathrm{star}} = 30^{\circ}$ (blue), $60^{\circ}$ (orange), and $90^{\circ}$ (green). The black and grey dashed lines show $\nu_0 = \Omega_{\mathrm{star}}$ and $\nu_1 = 2\Omega_{\mathrm{star}}$ respectively.}%
    \label{fig:psd}
\end{figure}
\autoref{fig:psd} shows the rms-normalized power spectrum densities (PSDs) of the observed bolometric luminosity and $\dot{M}$ time series shown in \autoref{fig:intensity-diffmag}(a) and (b).  For $\dot{M}$,  the PSDs in all the cases strictly follow $\nu^{-2}$  for all the frequencies. However, the power spectrum of luminosity shows a steeper trend in $\nu$ compared to $\dot{M}$. We observe that the time-variable signals also show considerable rms variability at the fundamental ($\nu_0$) and the first harmonic ($\nu_1$) frequencies. To compute the power within the peaks, we fit a Lorentzian function at $\nu_0$ and $\nu_1$ (\autoref{fig:psd}(a)) and integrate the power within the Lorentzian. The square root of the integral results in the following fractional rms variability amplitudes, $\rm rms_{\nu_0} = [0.054, 0.045]$ for $\chi_{\mathrm{star}} \in [30^{\circ}, 60^{\circ}]$, $\rm rms_{\nu_1} = [0.027, 0.073, 0.083]$  for $\chi_{\mathrm{star}} \in [30^{\circ}, 60^{\circ}, 90^{\circ}]$ respectively\footnote{There is no significant excess power above the accretion noise floor at $\nu_0$ for $\chi_{\rm star} = 90^{\circ}$ hence we do not compute the power.}. Thus, the fundamental decreases, and the first harmonic increases with increasing magnetic inclination. The non-pulsed broadband variability for all the cases is as follows: $\rm rms = [0.264, 0.159, 0.192]$ for $\chi_{\mathrm{star}} \in [30^{\circ}, 60^{\circ}, 90^{\circ}]$. Hence, broadband (accretion) noise dominates the overall lightcurve variability and is larger than the pulsed variability (i.e., the sum of fundamental and first harmonic)  by a factor of $1.3 - 3.3$.  The fractional rms of the peaks for the time variable signal closely follow the values in Section \ref{sec:avg_pulseprofiles} (see \autoref{fig:harmonic_content}). 
\par
\begin{figure}
    \centering
    \includegraphics[width=8.5cm, angle=0]{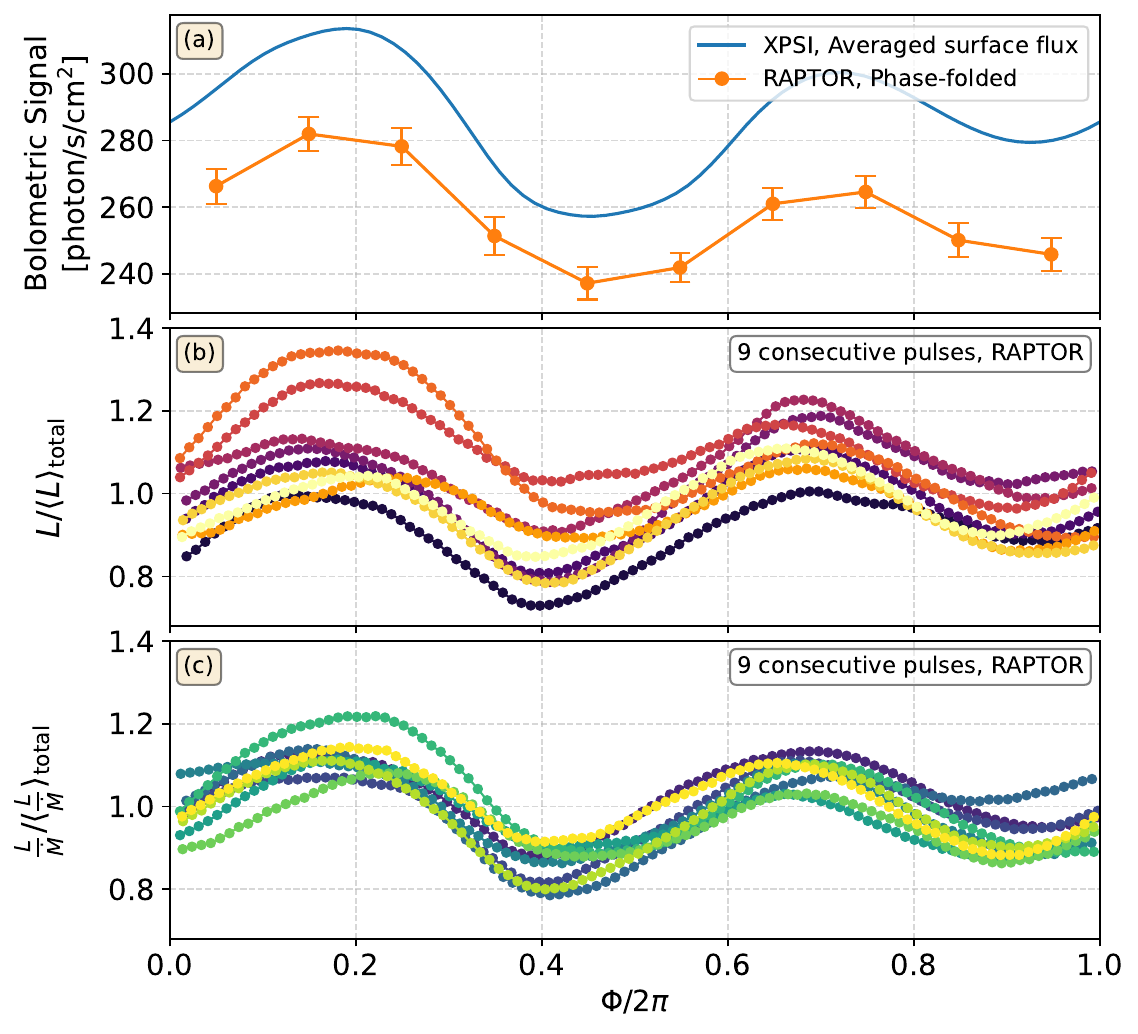}
    \caption{Pulse variability for $\chi_{\mathrm{star}} = 60^{\circ}$ with $\iobs = 70^{\circ}$. Panel (a) shows the variation between the pulse obtained from the averaged spots using $\xpsi$ and the phase-folded light curve from (\autoref{fig:intensity-diffmag}(a)) using $\raptor$. The vertical orange bars show the standard error in each phase bin. The middle and the lower panel show the pulse-to-pulse variations for 9 consecutive pulses obtained using $\raptor$. Panel (b) shows the pulse luminosity (L) normalized by the mean luminosity and panel (c) shows the $\dot{M}$ normalized equivalent of panel (b).}%
    \label{fig:pulses-variability}
\end{figure}

 
\autoref{fig:pulses-variability}(a) shows the phase folded pulse (orange) obtained from \autoref{fig:intensity-diffmag}(a) and the pulse from the time-averaged surface temperature map (blue curve) for $\chi_{\mathrm{star}} = 60^{\circ}$. The $\approx10\%$ difference in the pulsed lightcurves can be attributed to the non-linear nature of the operations: in essence, the phase folded bolometric lightcurve obtained with \raptor averages over the distribution of bolometric flux $\sim T^4$ compared to directly averaging the hotspot temperatures in the case of \xpsi. Further non-linearities are introduced by the raytracing procedures.  
\par
The variability in the light curve has two different aspects: (a) variations in the magnitude of matter-energy flux (and thus temperature) as a result of $\dot{M}$, (b) spot shape variability and temperature variability within the spot (highlighted by the movement of green contours in \autoref{fig:surfacemaps_10ms}). In order to separate the $\dot{M} $ variability from that imparted by the changing spot morphology, we detrend the light curve by normalizing it with $\dot{M}$ (\autoref{fig:intensity-diffmag} (c)). \autoref{fig:pulses-variability} (b, c) shows the pulse-to-pulse variation for 9 consecutive rotational cycles. Here, panel (b) shows the variation in the bolometric luminosity, and panel (c) shows the variation in $\dot{M}$ normalized bolometric luminosity. We observe that the variation in the $\dot{M}$ normalized case is notably smaller. 
To characterize the pulse-to-pulse variability, we fit a sine function (\autoref{eq:fit_harmonic}) to all the 47 rotational cycles shown in \autoref{fig:intensity-diffmag}. The coefficients of variation ($c_x$) \footnote{Here $x \in [B, C, \phi_1, \phi_2]$ for all the cycles.}  of the fundamental (B) and the first harmonic (C) amplitudes over 47 cycles are as follows: $c_B = 0.415$, $c_C = 0.152$ for $L/\dot{M}$ and $c_B = 0.601$, $c_C =0.251$ for the non-normalized light curve ($L$). $\phi_1$ and $\phi_2$ in \autoref{eq:fit_harmonic} characterize the pulse-to-pulse phase shifts of the fundamental and the first harmonic. The coefficients of variation for the phases ($\phi_1$ and $\phi_2$) are, $c_{\phi_1} = 0.814$ and $c_{\phi_2} = -0.279$ for $L/\dot{M}$ and $c_{\phi_1} = 0.911$, $c_{\phi_2} = -0.344$ for $L$. For the parameters chosen in this section, the variation in the fundamental is always larger compared to the first harmonic.
\subsection{Accretion Column}\label{sec:columns}
In this section, we study the impact of electron scattering in the plasma on the surface X-ray emission. To this end, we choose one rotational cycle and solve the radiative transfer equations along the geodesics, taking into account electron scattering absorption.

\begin{figure}
    \centering
    \includegraphics[width=8.5cm, angle=0]{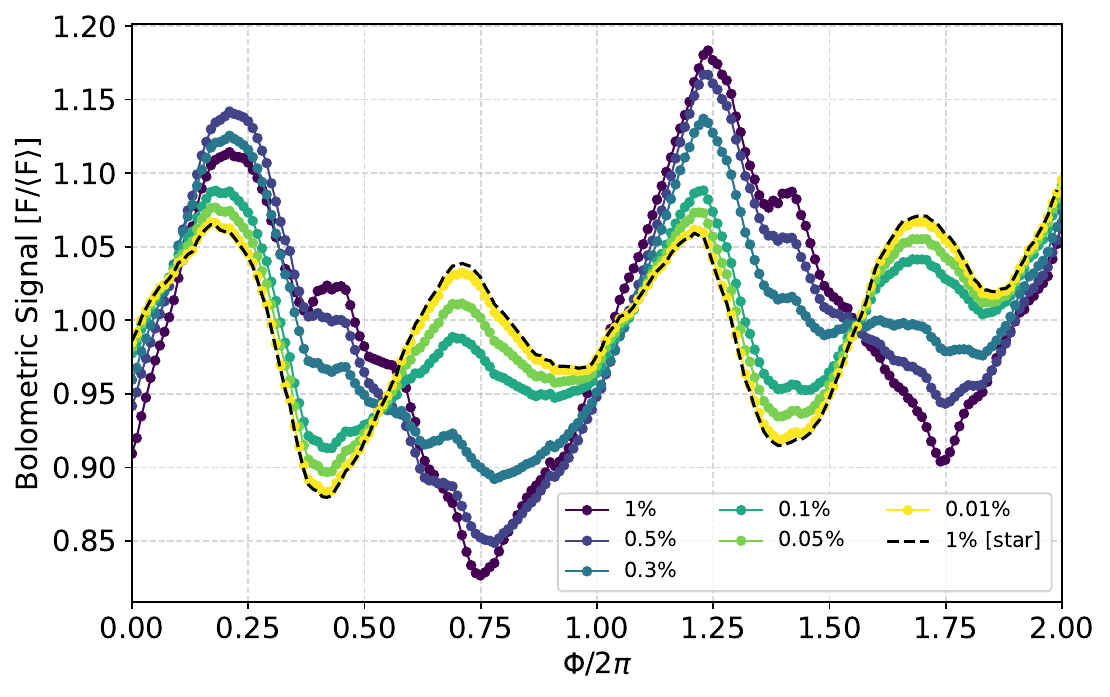}
    \caption{Variation in the pulses as a function of $\dot{M}$ for $\chi_{\rm star} = 60^{\circ}$ at \iobs $= 50^{\circ}$ using \raptor. Here, different colors represent the normalized bolometric flux for different mass accretion rates, and the black dashed line represents the pulse shape from the stellar black body without the electron scattering opacity.}%
    \label{fig:accretioncol_mdot}
\end{figure}

\autoref{fig:accretioncol_mdot} illustrates the variation in pulse shape as a function of mass accretion rate for $\chi_{\rm star} = 60^{\circ}$ and \iobs $= 50^{\circ}$. We begin with a mass accretion rate of $0.01\% \dot{M}_{\rm Edd}$ and gradually increase it to study the role of electron scattering absorption as the system transitions from quiescence to a moderate accretion rate. For a given GRMHD simulation, scaling the surface $\dot{M}$ to higher physical accretion rates is accompanied by an increased stellar magnetic field strength as given by Equation 6 in \cite{Das2022}. Following the unit conversion described in \autoref{sec:setup}, a mass accretion rate ranging from $0.01\% \dot{M}_{\rm Edd}$ to $1\% \dot{M}_{\rm Edd}$ corresponds to a magnetic field strength ranging from $\rm 3.78 \times 10^{7}\, G$ to $\rm 3.78 \times 10^{8}\, G$. This ensures a constant magnetospheric radius regardless of the physical mass accretion rate scaling. As we increase the mass accretion rate, two effects come into play: on the one hand, the surface emission increases; on the other hand, the optical depth also rises, which impacts the pulse profiles.

For $\chi_{\rm star} = 60^{\circ}$, there is one hotspot in the northern hemisphere and one in the southern hemisphere (antipodal dipolar geometry, see \autoref{fig:surfacemaps_avg}). For \iobs $= 50^{\circ}$, although the northern hotspot remains readily visible, the rays from the southern spot are obstructed by the accretion column and the disk. At $0.01\% \dot{M}_{\rm Edd}$, the absorption in the column is negligible, resulting in a pulse shape almost identical to the pulse from the surface emission alone. However, increasing $\dot{M}$ increases the absorption, resulting in a less visible southern spot and an increase in the fundamental amplitude. In particular, for the parameters adopted here, the pulse is completely dominated by the fundamental amplitude at $\approx 0.3\%\dot{M}_{\rm Edd}$. 
\begin{figure}
    \centering
        \includegraphics[width=8.5cm, angle=0]{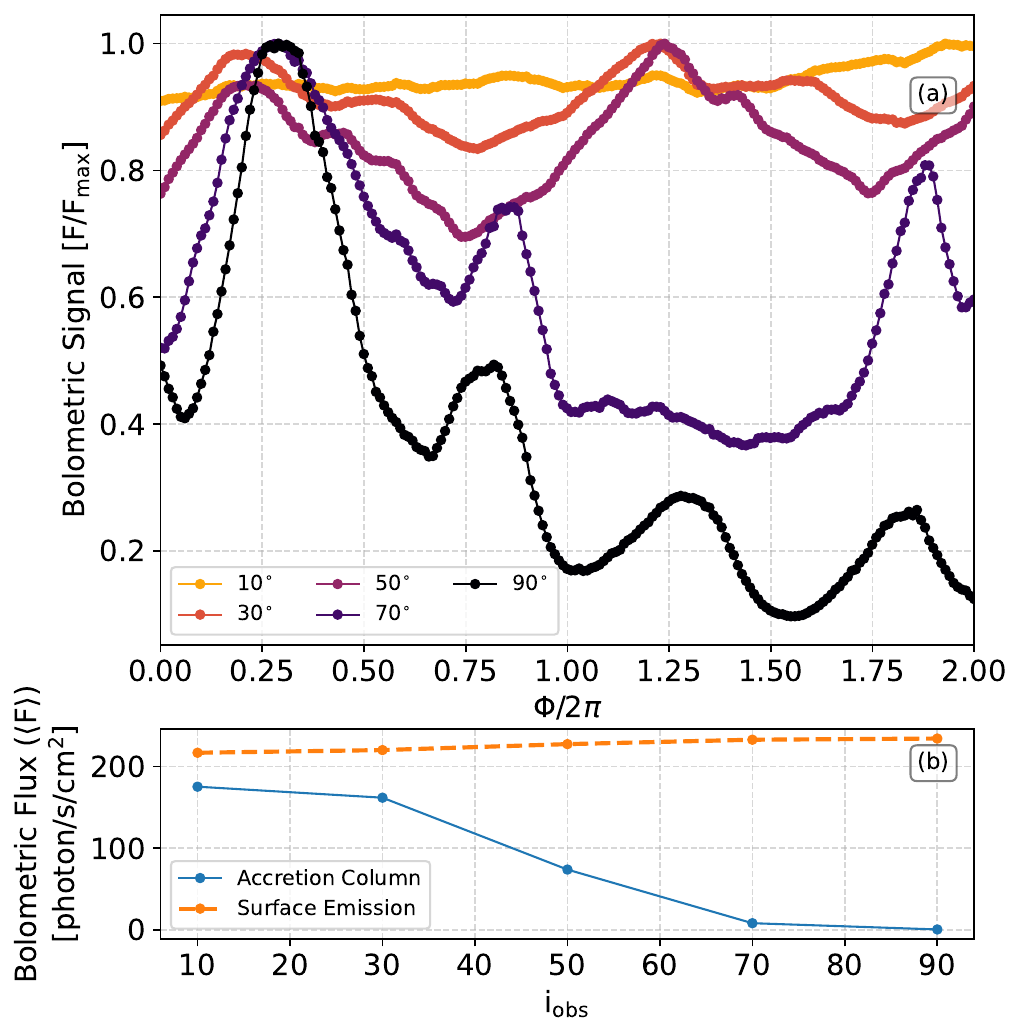}
    \caption{Variation in pulses as a function of observer inclination for $\chi_{\rm star} = 60^{\circ}$ for $\dot{M} = 1\% \dot{M}_{\rm Edd}$ using \raptor. Panel (a) and (b) show the variation in pulse shape and mean bolometric flux with \iobs\ respectively.}%
    \label{fig:accretioncol_iobs}
\end{figure}

\autoref{fig:accretioncol_iobs} shows the variation in the pulses as a function of observer inclination angle in the presence of accretion column and the disk at $1\%\dot{M}_{\rm Edd}$. Similar to the pulses shown in \autoref{fig:avgpp}, increasing \iobs\ results in larger harmonic content. However, taking absorption in the disk and the column obstructs the photons from the hotspot in the lower magnetic hemisphere resulting in a less dominant first harmonic with increasing \iobs\ compared to that observed in Section \ref{sec:avg_pulseprofiles}. Along with the variation in harmonic content, increasing \iobs\ also results in a decrease in average flux as shown in \autoref{fig:accretioncol_iobs} (b). We observe that for $\chi_{\rm star} = 60^{\circ}$, disk and column absorb $ \approx 50\%$ of the emitted surface flux for \iobs\ $\geq 50^{\circ}$. 
\par
The accretion disk results in a deviation of simple sinusoidal behavior of the antipodal spots, and increases the amplitudes of the higher harmonics.
For $\iobs = 50^{\circ}$ at a mass accretion rate of $1\% \dot{M}_{\rm Edd}$, the pulse originating from the stellar surface has fundamental and first harmonic fractional rms variability amplitudes of 0.067 and 0.054 respectively. The disk absorption results in the pulse having the fundamental, first and second harmonic fractional rms variability amplitudes of 0.113, 0.046 and 0.010 respectively. With increasing observer inclination, the impact of the equatorial accretion disk increases and results in more pronounced higher harmonics. While the surface X-ray pulse for $\iobs = 70^{\circ}$ has a fundamental and first harmonic fractional rms of 0.045 and 0.073, including absorption results in the fundamental, first, second, third and fourth harmonic fractional rms amplitudes of 0.193, 0.080, 0.061, 0.037, and 0.021 respectively. A figure depicting the impact of electron scattering absorption on the light curve can be found in \autoref{accretioncol_psds}.
\par
The electron scattering opacity of the turbulent accretion flow introduces significant additional variability in the light curve. The variability of the light curve, characterized in terms of the coefficient of variation, for \iobs $= 50^{\circ}$ increases from 0.18 to 0.213 once electron scattering absorption is taken into account. Increasing the observer inclination makes this effect even more pronounced. The variability in the light curve at \iobs $= 70^{\circ}$ from the surface emission alone is observed to be 0.182, and increases to 0.391 once absorption is considered. 

\section{Discussion $\&$ Conclusions} \label{sec:discussion}
In this paper, we have analyzed the accretion hotspot properties using GRMHD simulations for rotating magnetized stars with a range of stellar magnetic inclinations $\chi_{\rm star} \in [30^{\circ}, 60^{\circ}, 90^{\circ}]$. The stellar magnetic field strength and angular frequency for all the runs studied here are set to be in the range observed for the accreting millisecond pulsars.
\par
The geometric properties of the averaged hotspots can be summarized as follows:
for the smallest inclination considered here $\chi_{\rm star} = 30^{\circ}$, the averaged spot shapes resemble crescents around the magnetic axis. As we increase $\chi_{\rm star}$, the crescents are transformed into bands below and above the magnetic axis in the upper and lower magnetic hemispheres respectively. MHD simulations of accreting stars in a Newtonian and pseudo-Newtonian potential have been previously employed to study the hotspot properties at the stellar surface \citep{Romanova2004, Kulkarni2005}, and the overall hotspot shapes in our simulations are roughly consistent with those studies. Most of the detailed studies of the accreting hotspots in the non-relativistic regime were performed for non-magnetized disks. Some simulations of stars with MRI-driven accretion disks focusing on magnetospheric interactions have been previously performed, but, only for lower magnetic inclinations ($\chi_{\rm star} \lesssim 30^{\circ}$), and show near ring-like hotspots around the rotation axis \citep{Romanova2012}. 
\par
Although complex spot shapes involving, e.g., crescents have been applied in pulse profile modeling studies of rotation-powered pulsars \citep{Riley2019}, to date, PPM studies for accretion powered pulsars have been adopting circular hotspot geometries \citep[see e.g.][]{Salmi18, Salmi2021, Ahlberg2024, Dorsman2025}. The family of crescent and elongated band shaped hotspots resulting from the global GRMHD simulations, along with the impact these complex spot geometries have on the simulated pulse profiles, strongly suggests the need to move beyond the assumption of circular spots in case of accretion-powered pulsars. 
\par
We also observe significant variability of the shapes and temperatures of the hotspots in all our simulations. In contrast to the Rayleigh-Taylor induced variability in the previous $\alpha$-disk simulations \citep{Romanova2004, Kulkarni2005}, in our case, the hotspot variability stems from both the MRI and the Rayleigh-Taylor instabilities resulting in variations on a timescale similar to that of the stellar rotational period. 
\par
As the star rotates, the non-axisymmetric hotspots give rise to X-ray pulsations. To quantify the expected pulse properties, as a first step, using relativistic ray tracing, we have simulated the resulting pulsed X-ray signal resulting from the time-averaged accretion spots. For all the magnetic inclinations explored here ($\chi_{\rm star} \in [30^{\circ}, 60^{\circ}, 90^{\circ}]$), we find that the fractional rms amplitudes of the fundamental and the first harmonic usually range between $1 - 12\%$. For the known population of AMXPs including transitional pulsars, observed fractional amplitude in the peak of an outburst is typically in the range 3-10\% rms \citep{Wijnands98,Markwardt02,Galloway02,Campana03,Strohmayer03,Galloway05,Altamirano10,Altamirano11,Papitto11,Ferrigno14,Papitto15,Strohmayer17,Bult18,Sanna18b,Papitto19,Ng21,Bult22,Sanna22}, very occasionally higher \citep{Riggio11,Sanna18,Bult19}. There is also a population of intermittent AMXPs, for which the amplitudes of the oscillations - when detected - are at the lower end of this range \citep{Kaaret06,Altamirano08,Casella08}, close to the detectability limit.  Many but not all AMXPs have harmonic content in the form of a detectable first harmonic: while for most sources the fundamental is stronger than the first harmonic, there are some AMXPs where the first harmonic is stronger \citep{Altamirano11,Papitto15}. The ranges of amplitudes found for the cases simulated in our study are consistent with the observed values. Provided that magnetic inclination is not too high, the amplitude of the fundamental in our simulations is generally higher than that of the first harmonic, except for high observer inclinations - which would be consistent with a dominant harmonic being less common. 
\par 
The turbulent accretion flow imparts significant time variability to the accretion hotspots and, thus, the corresponding surface X-ray emission. Ray tracing different snapshots in our GRMHD simulations allows us to study this time variability and how it is reflected in the pulse profiles. We observe that in the absence of any obstructions by the accretion disk, the X-ray light curves strongly correlate with the corresponding time evolution of the surface mass accretion rate. There is significant variability in all the simulated light curves introduced by two factors: (a) stellar rotation and (b) the turbulent accretion flow. In our simulations, the overall light-curve variance is dominated by the broadband accretion noise which is larger than the pulse variability by a factor of $1.3 - 3.3$. Consistent with the case of a fixed-in-time hotspot temperature profile, for the time-variable signal, the fractional rms amplitudes of the fundamental decreases and the first harmonic increases with increasing $\chi_{\rm star}$. 
\par
The presence of the accretion flow also leads to a significant pulse-to-pulse variation, which can be attributed to variations in the magnitude of matter-energy flux (and thus temperature) due to $\dot{M}$, and spot shape variability (due to varying magnetospheric coupling). For a fiducial run with $\chi_{\rm star} = 60^{\circ}$ at \iobs $= 70^{\circ}$, the intrinsic pulse-to-pulse variation characterized in terms of coefficient of variation of the pulse amplitudes and phases is observed to be 0.601 (fundamental amplitude), 0.251 (first harmonic amplitude), 0.911 (fundamental phase), and 0.342 (first harmonic phase). Since the average spot temperature strongly correlates with $\dot{M}$, the variations in the light curve due to the variable matter-energy flux magnitude and spot shape variability can be separated, and we observe that even the variation of the hotspot structures alone can impart a significant pulse-to-pulse variability. The variability index resulting from the spot shape fluctuations is found to be 0.415 for the fundamental amplitude, 0.152 for the first harmonic amplitude, 0.814 for the fundamental phase, and 0.279 for the first harmonic phase. We always observe a larger variation in the fundamental compared to the first harmonic. 
\par
Our simulations indicate that the pulse properties generated from the global 3D GRMHD simulations change on timescales similar to the rotation period. Even though current observations cannot directly measure the accretion variability in as short time scales as studied here, it is important to study how parameter recovery via pulse profile modeling would be affected by ignoring the variations. Similar studies for thermonuclear burst oscillation sources have shown that biased parameter constraints are possible even if the time-averaged pulse profile seems roughly constant \citep{Kini2023,Kini2024,Kini2024b}.
\par 
In the case of accreting neutron stars, the X-ray pulses produced from the stellar surface is significantly influenced by the plasma in the accretion column and the disk. Simple analytical models of stars with single accretion hotspots have demonstrated that electron scattering absorption plays a key role in determining the pulse characteristics \citep{Ahlberg2024}. In this paper we have investigated the role of electron scattering opacity of the turbulent accretion flow on the pulse characteristics of a star with dipolar magnetic field for a range of persistent mass accretion rates and observer inclinations. We observe that both the pulse phase and amplitudes depend significantly on the mass accretion rate. With increasing mass accretion rate, on one hand the surface accretion hotspots become hotter and, on the other hand, the disk absorption increases and leads to an increasing obscuration of the southern hotspot. 
\par
In a physical system, a decreasing mass accretion rate is accompanied with a receding accretion disk, and has been proposed as a plausible cause of visible higher harmonics in the pulse \citep{Ibragimov2009}. In our scale-free GRMHD simulations,  we vary the accretion rate by scaling the same simulation to different physical units. As shown in \cite{Das2022}, the $\dot{M}$ scaling determines the stellar magnetic field strength and thus a constant magnetospheric radius is obtained regardless of the physical $\dot{M}$ values. We observe that while an increasing mass accretion rate increases (reduces) the fundamental (first harmonic) fractional rms due to reduced visibility of the southern hotspot, it also introduces higher harmonics in the light curve. Our simulations show that taking absorption into account also results in more variable bolometric light curves. For example, at $1\% \dot{M}_{\rm Edd}$, the variability of the light curve increases from 0.18 to 0.213 in presence of absorption at $\iobs = 50^{\circ}$, and from 0.182 to 0.391 at an observer inclination of $70^{\circ}$.
\par
Overall, the impact of electron scattering absorption in the light curve is two fold: one, it introduces significant additional variability, and, second, it generates higher harmonics in the bolometric light curve. Due to the equatorial disk, increasing the observer inclination enhances both of these effects and results in more pronounced higher harmonics and variability.
\par
Our current study treats scattering only as (monochromatic) absorption opacity which dims the signal from the pulse but ignores line-of-sight scattering and energy scattering. 
We have therefore focused on results regarding the bolometric pulse properties. 
Including scattering into the line of sight would predominantly affect the southern hotspot which is strongly absorbed by the accretion disk matter starting at Eddington fractions of $1\%$.  Due to directional scattering, we thus expect the dips of the first harmonic to be less pronounced and broadened compared to the case studied here.  
Furthermore, we note that our predicted pulse amplitudes can be slightly overestimated since we treat the initial hotspot spectrum and beaming as isotropic blackbody here (see e.g. Figure 4 of \citealt{Poutanen2003} for the effect). More accurate modeling of the neutron star atmosphere, subsequent scattering, as well as incorporating the background emission from the disk are necessary to make predictions on the pulse properties with energy and assess the effect of line-of-sight scattering.  In particular, determining the disk emission `background' is an important task for mass-radius inference of such sources \citep[][]{Patruno2009a, Kini2024b, Dorsman2025}.   This requires more advanced radiation transfer calculations which we leave for future work.  Furthermore, it will be very interesting to sample pulse profiles for a number of simulations with varying magnetospheric radii.  This will allow us to model changes in pulse profiles with varying mass accretion rate onto a given source \citep[e.g.][]{Hartman2008, Patruno2009, Patruno2010}.

\begin{acknowledgments}
We thank Joonas N\"attil\"a for valuable inputs on code comparison. We would also like to thank Phil Uttley for interesting discussions on variability.  
\par
PD and OP acknowledge funding from the Virtual Institute for
Accretion (VIA) within NOVA (Nederlandse Onderzoeksschool
voor Astronomie) Network 3 ‘Astrophysics in extreme conditions’.
Simulations have been carried out in part on the HELIOS cluster of
the Anton Pannekoek Institute for Astronomy and on the Dutch national e-infrastructure with the support of SURF Cooperative. TS and ALW
acknowledge support from ERC Consolidator Grant No. 865768
AEONS (PI Watts).
\par
JD is supported by NASA through the NASA Hubble Fellowship grant HST-HF2-51552.001A, awarded by the Space Telescope Science Institute, which is operated by the Association of Universities for Research in Astronomy, Incorporated, under NASA contract NAS5-26555.
\end{acknowledgments}
%

\vspace{5mm}
\facilities{\textit{Snellius, Dardel}}


\software{\bhac \citep{Porth2017},
            \raptor \citep{Bronzwaer2018, Bronzwaer2020},
            \xpsi \citep{xpsi},
          Python \citep{Python2007},
          NumPy \citep{Numpy2011},
          Scipy \citep{Jones2001},
          MPI \citep{mpi4py},
          Matplotlib \citep{Hunter2007, matplotlibv2},
          Mayavi \citep{Mayavi},
          IPython \citep{IPython2007}}


\bibliography{References,References-new,amxprefs}{}
\bibliographystyle{aasjournal}

\appendix
\section{Accretion Columns}\label{volume_rendering}
\begin{figure*}
    \centering
    \parbox[b]{0.48\textwidth}{
        \centering
        \textbf{$\chi_{\rm star} = 60^{\circ}$} \\[0.5ex]
    \includegraphics[width=8.9cm, angle=0]{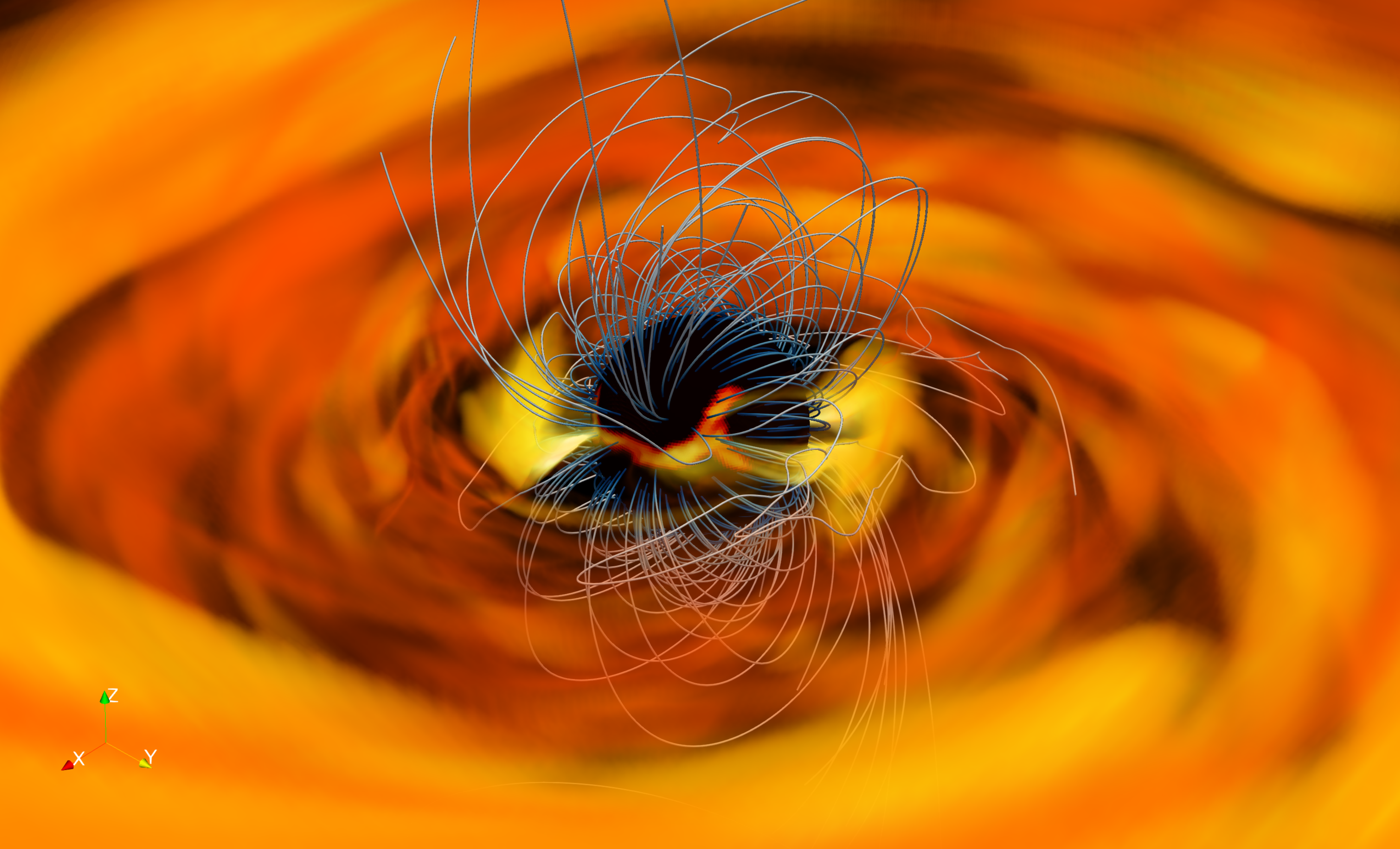}}
    \hfill
    \parbox[b]{0.48\textwidth}{
        \centering
        \textbf{$\chi_{\rm star} = 90^{\circ}$} \\[0.5ex]
    \includegraphics[width=8.9cm, angle=0]{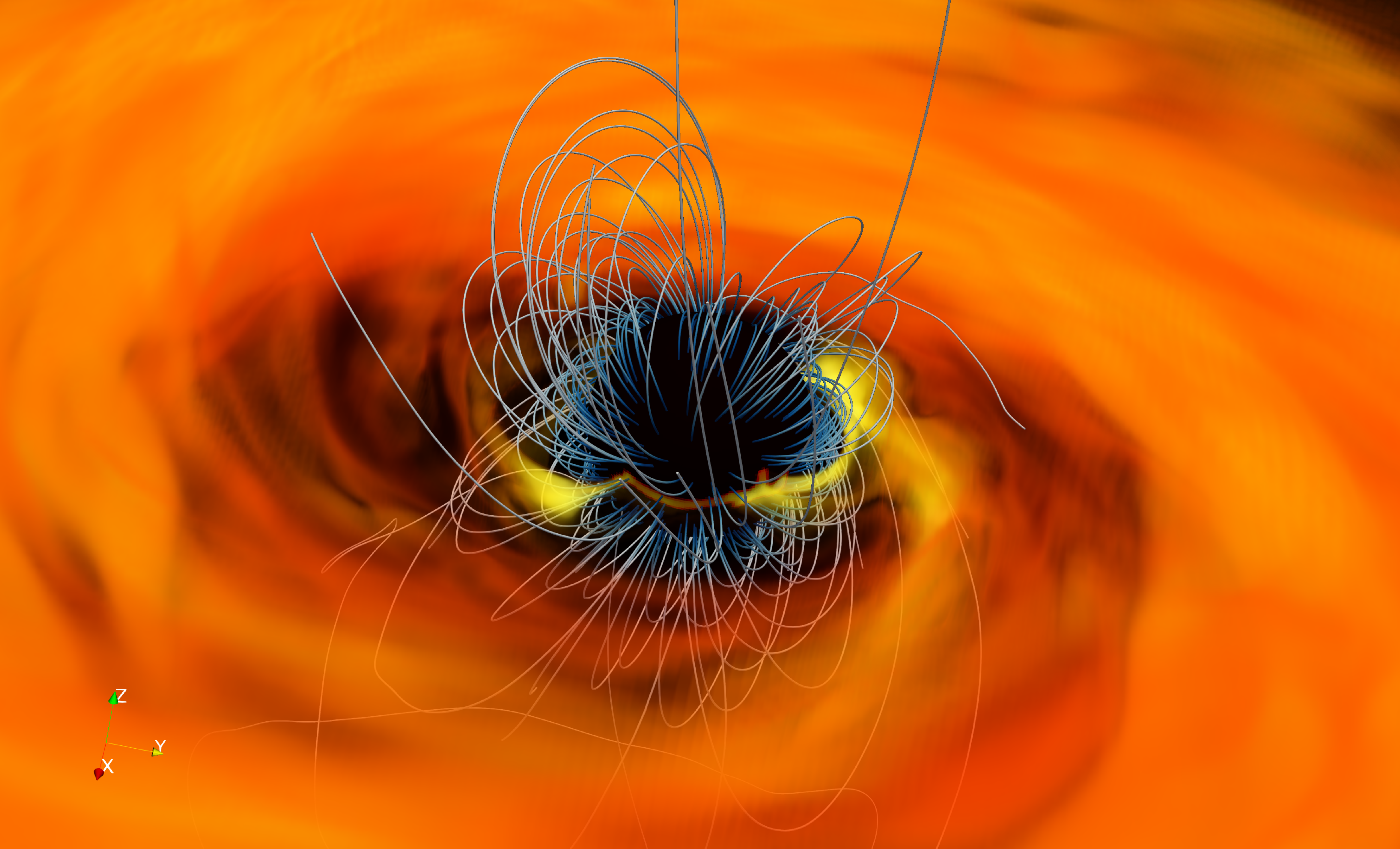}}
    \caption{Volume rendering of density for $\chi_{\rm star} = 60^{\circ}$ (left panel) and $\chi_{\rm star} = 90^{\circ}$ (right panel). The solid lines show the corresponding stellar magnetic fieldlines for each case.}
    \label{fig:accretioncolumn}
\end{figure*}

\autoref{fig:accretioncolumn} shows the volume rendering of density for $\chi_{\rm star} = 60^{\circ}$ (top panel) and $\chi_{\rm star} = 90^{\circ}$ (bottom panel). The solid lines show the magnetic field, and the color gradient in the lines shows the magnitude of the field. Both panels clearly show the formation of symmetric accretion columns across the magnetic equator (magnetic axis in the $z-x$ plane, axes shown in the lower left corner of the plot) close to the $\mu-y$ plane. The sphere at the center of the plots shows the density map at the stellar surface. 
\section{Power Spectrum Densities of the bolometric light curve with absorption}\label{accretioncol_psds}
\begin{figure*}
    \centering
    \includegraphics[width=7.8cm, angle=0]{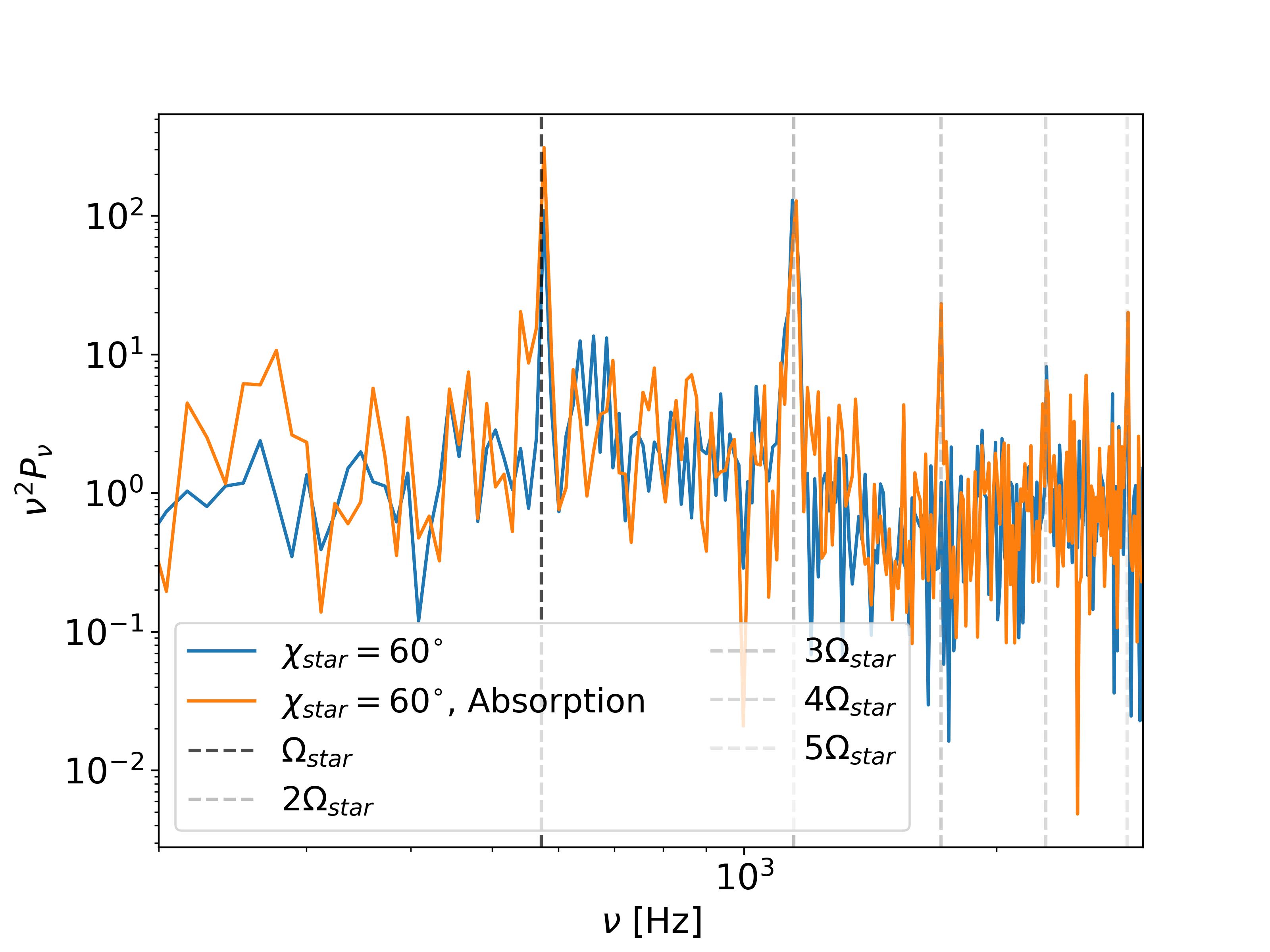}%
    \includegraphics[width=7.8cm, angle=0]{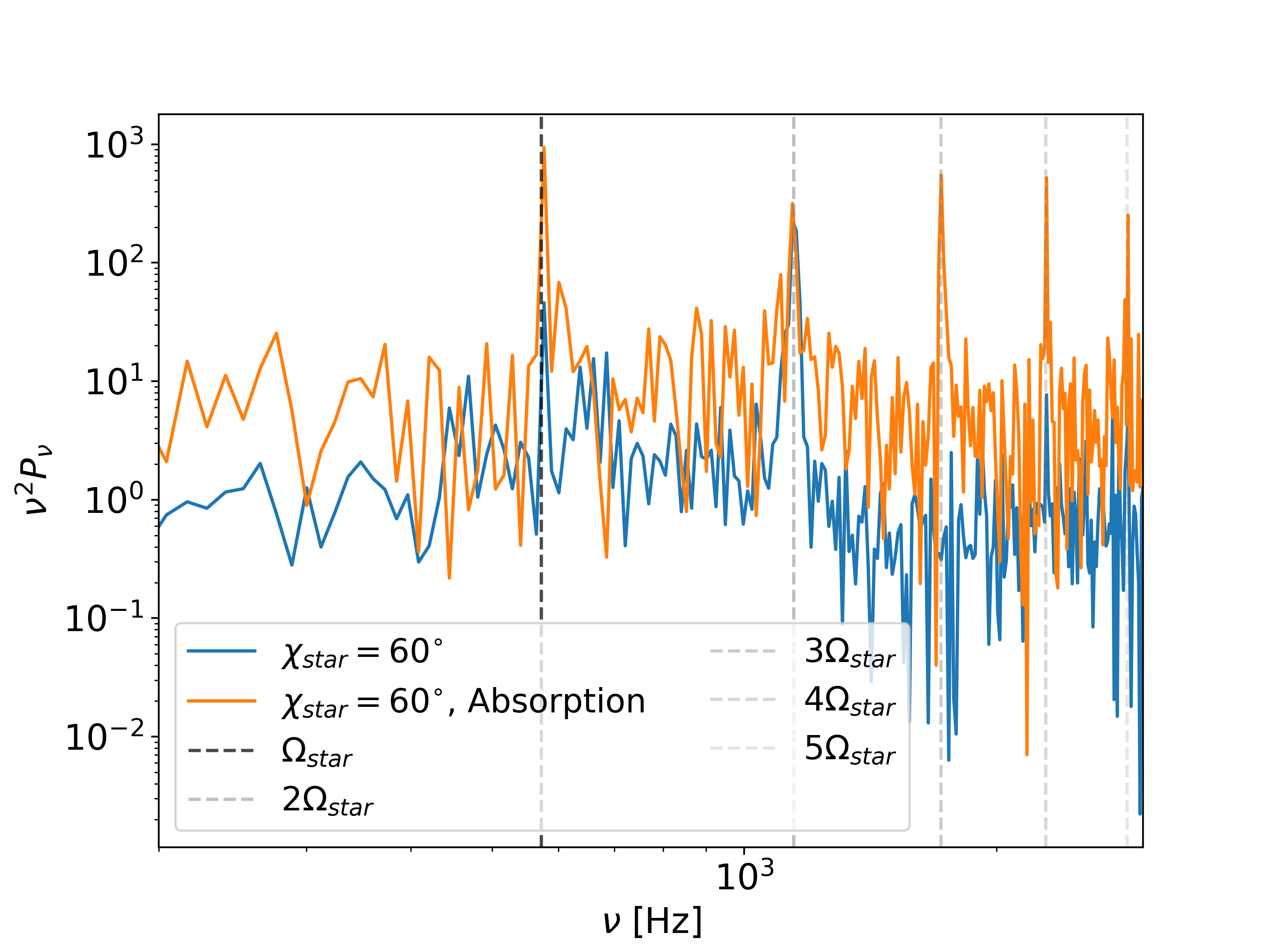}
    \caption{RMS-normalized PSDs of the bolometric luminosity with electron scattering absorption for different observer inclinations, $\iobs = 50^{\circ}$ (left) and $\iobs = 70^{\circ}$ (right) for $\chi_{\rm star} = 60^{\circ}$ at $1\%\dot{M}_{\rm Edd}$. The blue and orange curves represent the PSDs resulting from the surface hotspots alone and surface emission with electron scattering absorption respectively. The black and grey dashed lines, with decreasing opacity, indicate $\nu_0 = \Omega_{\mathrm{star}}$, $\nu_1 = 2\Omega_{\mathrm{star}}$, $\nu_2 = 3\Omega_{\mathrm{star}}$, $\nu_3 = 4\Omega_{\mathrm{star}}$ and $\nu_4 = 5\Omega_{\mathrm{star}}$ respectively.}
    \label{fig:PSDcolumn}
\end{figure*}

\autoref{fig:PSDcolumn} illustrates the rms-normalized power spectrum densities of the bolometric luminosity with electron scattering absorption for different observer inclinations for $\chi_{\rm star} = 60^{\circ}$. The accretion column and the disk introduces significant additional variability and higher harmonics in the light curve. Since we have an equatorial accretion disk, this effect is even more evident for higher observer inclinations.




\end{document}